\documentclass[twocolumn,prl,amsmath,amssymb,superscriptaddress]{revtex4-1}
\usepackage{graphicx,dsfont,color}
\usepackage{bm}
\usepackage{bbold}
\usepackage[colorlinks=true,citecolor=blue,urlcolor=blue]{hyperref}
\begin{document}

\title{Non-Hermitian Waveguide Cavity QED with Tunable Atomic Mirrors}
\author{Wei Nie}\email{weinie@tju.edu.cn}
\affiliation{Center for Joint Quantum Studies and Department of Physics, School of Science, Tianjin University, Tianjin 300350, China}
\author{Tao Shi}
\affiliation{Institute of Theoretical Physics, Chinese Academy of Sciences, P.O.
Box 2735, Beijing 100190, China}
\affiliation{CAS Center for Excellence in Topological Quantum Computation,
University of Chinese Academy of Sciences, Beijing 100049, China}
\author{Yu-xi Liu}
\affiliation{School of Integrated Circuits, Tsinghua University, Beijing 100084, China}
\author{Franco Nori}
\affiliation{Theoretical Quantum Physics Laboratory, Cluster for Pioneering Research, RIKEN, Wakoshi, Saitama 351-0198, Japan}
\affiliation{Center for Quantum Computing, RIKEN, Wakoshi, Saitama 351-0198, Japan}
\affiliation{Physics Department, The University of Michigan, Ann Arbor, Michigan 48109-1040, USA}

\begin{abstract}
Optical mirrors determine cavity properties by means of light reflection. Imperfect reflection gives rise to open cavities with photon loss. We study an open cavity made of atom-dimer mirrors with a tunable reflection spectrum. We find that the atomic cavity shows anti-$\mathcal{PT}$ symmetry. The anti-$\mathcal{PT}$ phase transition controlled by atomic couplings in mirrors indicates the emergence of two degenerate cavity supermodes. Interestingly, a threshold of mirror reflection is identified for realizing strong coherent cavity-atom coupling. This reflection threshold reveals the criterion of atomic mirrors to produce a good cavity. Moreover, cavity quantum electrodynamics with a probe atom shows mirror-tuned properties, including reflection-dependent polaritons formed by the cavity and probe atom. Our work presents a non-Hermitian theory of an anti-$\mathcal{PT}$ atomic cavity, which may have applications in quantum optics and quantum computation.
\end{abstract}

\maketitle

\textit{Introduction}.---Quantum cavities are the cornerstone of quantum optics for interfacing light-matter interaction~\cite{scully1997quantum,Agarwal2012,Haroche2006book,RevModPhys.85.299}. The Fabry-P\'{e}rot interferometer is a cavity that consists of two parallel optical mirrors. A Fabry-P\'{e}rot cavity can be realized in many quantum systems with varieties of light reflectors~\cite{kavokin2017}. Meanwhile, Fabry-P\'{e}rot cavities can be synthesized with quantum mirrors, e.g., atoms or artificial atoms~\cite{PhysRevA.78.063827,PhysRevA.83.013825,PhysRevLett.107.133002,Chang2012cavity,PhysRevLett.113.243601,hoi2015probing,PhysRevLett.123.233602}, based on the remarkable photon scattering effects of single atoms in one-dimensional (1D) waveguides~\cite{PhysRevA.65.033832,PhysRevLett.95.213001,chang2007single,PhysRevLett.101.100501,PhysRevA.80.062109,astafiev2010resonance,peropadre2013scattering,van2013photon,
lodahl2017chiral}. Recently, experiments have realized strong coupling in cavity quantum electrodynamics (QED) with single-atom mirrors~\cite{mirhosseini2019cavity}. Large reflection of atomic mirrors at the central frequency gives rise to a dark mode, i.e., the effective cavity mode~\cite{PhysRevA.78.063827,Chang2012cavity}. Atomic cavities receive much attention for studying quantum optics with tailored light-matter interaction~\cite{PhysRevA.78.053806,PhysRevA.79.063847,
PhysRevA.93.023808,PhysRevLett.122.073601,song2021optical,PhysRevA.104.L031701,PhysRevA.105.033705}. However, due to the limitation in tuning mirror reflection, atomic cavities are far from being thoroughly understood.

Generally speaking, photon loss is unavoidable in quantum cavities~\cite{dutra2005cavity}, because of imperfect mirror reflections that lead to the coupling between cavity modes and continuum in free space~\cite{PhysRevLett.82.3787,lalanne2008photon,PhysRevB.87.115419,PhysRevA.92.053810}. Indeed, open cavities have substantial applications in quantum computation~\cite{PhysRevLett.75.3788,gu2017microwave} and quantum networks~\cite{kimble2008quantum,RevModPhys.87.1379,PhysRevApplied.15.054043}. Knowing how mirrors alter photon reflections is central to design novel quantum devices~\cite{RevModPhys.86.1391,PhysRevA.92.063836,qian2021spontaneous}. For example, tailored reflection (transmission) of cavities is useful for practical quantum technologies, such as single-photon resources~\cite{Friedler2009solid,PhysRevB.98.121306} and Fano lasers~\cite{PhysRevLett.113.163901,yu2017demonstration}. A puzzle naturally arises: what is the fate of an open cavity if the mirror reflection is strongly modified? Different from previous theories of open cavities based on conventional light reflectors~\cite{PhysRevLett.82.3787,lalanne2008photon,PhysRevB.87.115419,PhysRevA.92.053810,Friedler2009solid,PhysRevB.98.121306,PhysRevLett.113.163901,
yu2017demonstration}, waveguide-interfaced quantum light-matter interactions~\cite{RevModPhys.90.031002,RevModPhys.95.015002} make atomic cavities an excellent platform to study cavity QED~\cite{PhysRevA.78.063827,PhysRevA.83.013825,PhysRevLett.107.133002,Chang2012cavity,PhysRevLett.113.243601,hoi2015probing,PhysRevLett.123.233602,mirhosseini2019cavity,
PhysRevA.78.053806,PhysRevA.79.063847,PhysRevA.93.023808,PhysRevLett.122.073601,song2021optical,PhysRevA.104.L031701,PhysRevA.105.033705}, and might shed new light on open cavities.

\begin{figure}[b]
\includegraphics[width=8.5cm]{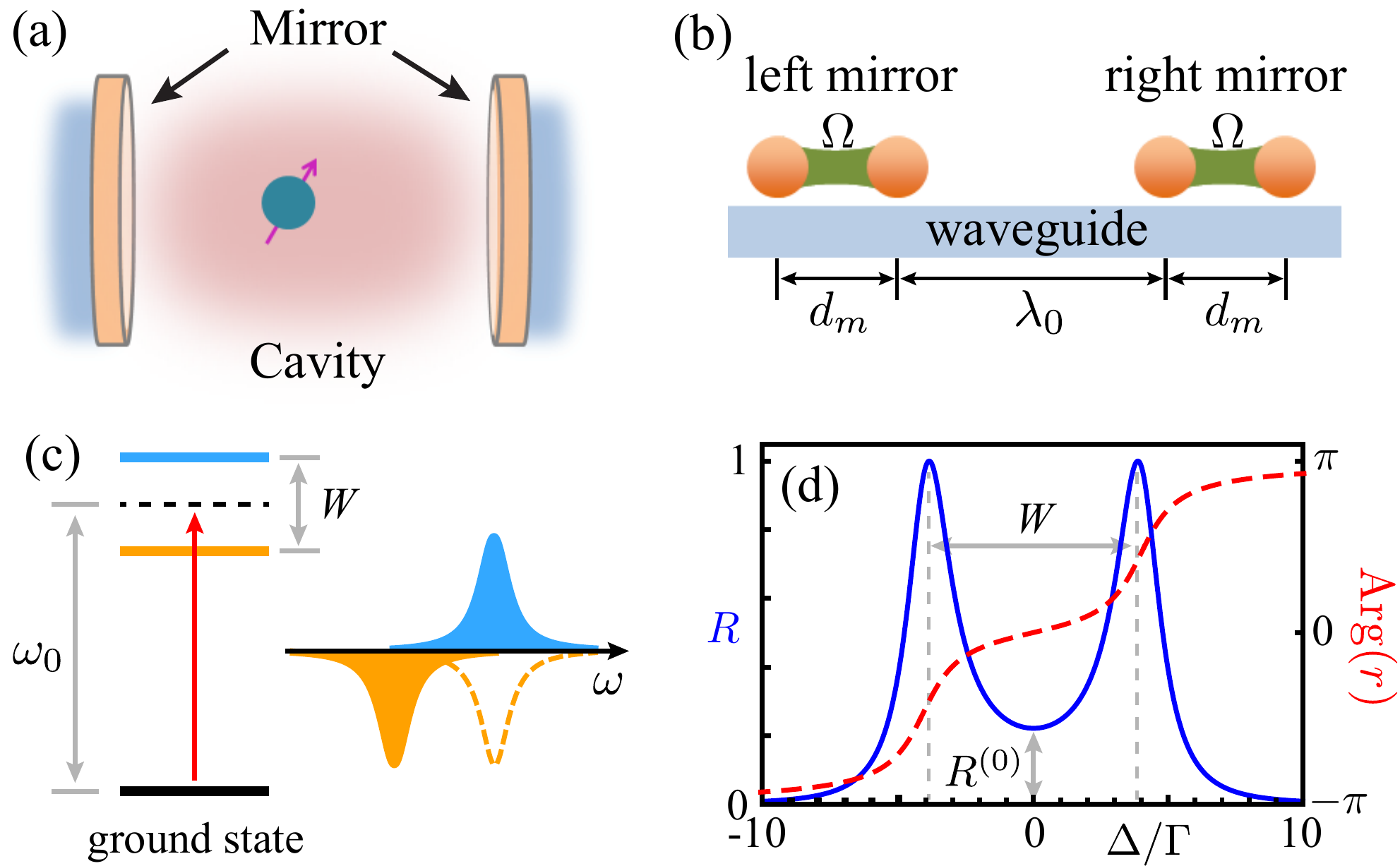}
\caption{(a) Schematic of a Fabry-P\'{e}rot cavity with two parallel mirrors. (b) Atomic cavity consisting of two atom-dimer mirrors, separated by a distance $\lambda_0$, coupled to a waveguide. For each mirror, we consider a direct atomic coupling $\Omega$ and anti-Bragg condition $d_m=\lambda_0/4$. (c) Energy levels of scattering states in the anti-Bragg atom-dimer mirror with $W=2(\Omega + \Gamma)$ (left), and corresponding reflected photon amplitudes (right). (d) Reflection $R=|r|^2$ (blue-solid) and photon phase shift $\mathrm{Arg}(r)$ (red-dashed) produced by the atom-dimer mirror.}\label{cavity}
\end{figure}

In this work, we study a cavity consisting of atomic mirrors with two-coupled atoms, i.e., atom dimer, coupled to a 1D coplanar waveguide in superconducting quantum circuits~\cite{kannan2023}.  We find that the atom-dimer cavity has anti-parity-time (anti-$\mathcal{PT}$) symmetry. Different from previous works that tune anti-$\mathcal{PT}$ symmetric systems via frequency detunings~\cite{peng2016anti,choi2018observation,li2019anti,PhysRevLett.123.193604,PhysRevApplied.13.014053,zhang2020breaking,PhysRevLett.125.147202,PhysRevLett.126.180401,
yang2022radiative}, here atomic couplings in mirrors alter light reflection and induce non-Hermitian phase transitions of the cavity. Therefore, two degenerate cavity supermodes are created. We propose a non-Hermitian theory for atomic cavity QED.

\textit{Tunable reflection by anti-Bragg atom-dimer mirror}.---Figure~\ref{cavity}(a) shows the schematics of a Fabry-P\'{e}rot cavity with a probe atom. The mirror can be realized, e.g., with a single two-level atom coupled to a waveguide~\cite{PhysRevLett.95.213001,chang2007single,PhysRevLett.101.100501}. The single-atom reflection spectrum has a Lorentzian line shape~\cite{PhysRevLett.95.213001} $R(\Delta)= |\Gamma/(\Delta + i\Gamma)|^2$, with atomic decay rate $\Gamma$ and detuning $\Delta=\omega-\omega_0$. Here, $\omega$ and $\omega_0$ are the frequencies of the driving field and atom, respectively. Due to this single-peak reflection, a dark cavity mode can be generated in an atomic cavity~\cite{mirhosseini2019cavity}.

Two atoms produce a scattering state with radiation rate $2\Gamma$~\cite{van2013photon,tiranov2023} for an atomic spacing $n\lambda_0/4$, with an even number $n$ and single-photon wavelength $\lambda_0=2\pi c/\omega_0$, i.e., the Bragg condition~\cite{PhysRevLett.117.133603,PhysRevLett.117.133604}. Such collectively enhanced scattering state (superradiant state) leads to broad-band reflection with a Lorentzian profile, useful for high-finesse cavities~\cite{Chang2012cavity}. However, the scenario is different for spacing $m\lambda_0/4$ (with $m$ an odd number), i.e., the anti-Bragg condition~\cite{Huebner1996,PhysRevA.106.L031702}. We study an atom dimer with $d_m=\lambda_0/4$~\cite{kannan2023}, as shown in Fig.~\ref{cavity}(b). The master equation is ($\hbar = 1$) $\dot{\rho}(t)=-i[H_m,\rho(t)] + \mathcal{D}[\rho]$, where the Hamiltonian is $H_m = \sum_{j=1,2} \omega_0\sigma_j^+ \sigma_j^- + (\Omega + \Gamma) (\sigma_1^+ \sigma_2^- + \mathrm{H.c.})$, with a direct atomic coupling $\Omega$ and waveguide-mediated dispersive coupling $\Gamma$. Here, $\sigma_{j}^+ = |e_j\rangle \langle g_j|$ with the ground (excited) state $|g\rangle$ ($|e\rangle$). The Lindblad operator is $\mathcal{D}[\rho]=\sum_{j} \Gamma (2 \sigma_{j}^- \rho \sigma_{j}^+ - \sigma_{j}^+ \sigma_{j}^- \rho - \rho \sigma_{j}^+ \sigma_{j}^-)$. In this mirror, there are two scattering states $\Phi_{\pm} =(1/\sqrt{2})(1,\pm 1)^{\top}$ with equal decay rate $\Gamma$, as shown in Fig.~\ref{cavity}(c). The photon amplitude reflected by the anti-Bragg atom-dimer mirror is~\cite{SupplementalMaterial}
\begin{equation}\label{EqR}
r(\Delta)= \sum_{n=\pm} (-1)^{n} \frac{\Gamma}{(\Delta - \delta_n + i \Gamma)},
\end{equation}
where $\delta_{\pm}$ are frequencies of the scattering states with respect to $\omega_0$. Quantum interference between reflected photons is determined by the frequency difference $W=2(\Omega + \Gamma)$ between two scattering states. Figure~\ref{cavity}(d) shows the intensity and phase shift of the reflected photon. At the central frequency $\Delta=0$, phase shifts $0$ and $\pi$ are produced, respectively, for anti-Bragg atom-dimer mirrors with $W>0$ and $W<0$, due to swapping of scattering states. Without loss of generality, we focus on the regime $W\geq0$.

\begin{figure}[t]
\includegraphics[width=8.5cm]{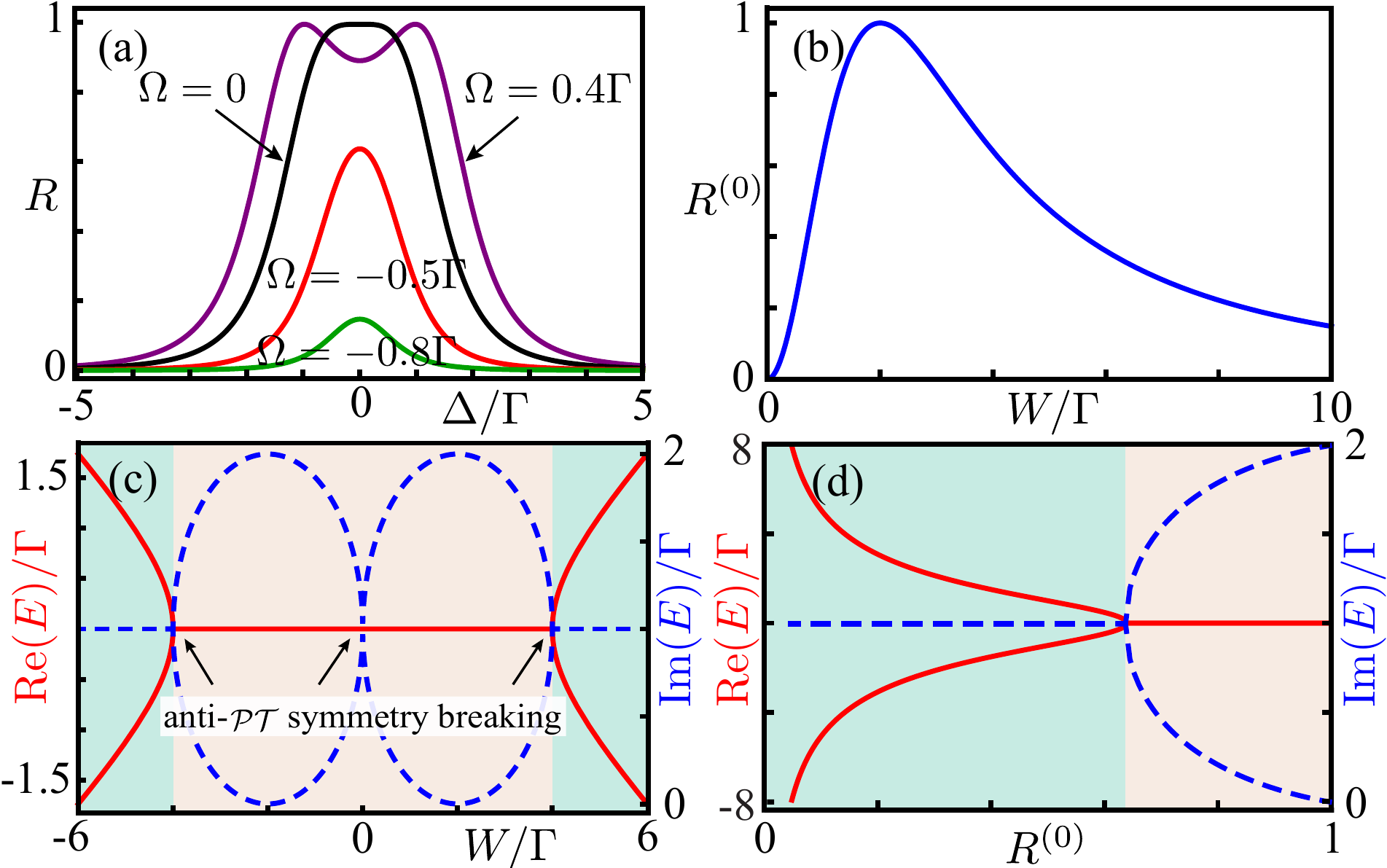}
\caption{(a) Reflection of the atom-dimer mirror for various values of atomic coupling. (b) Reflection $R^{(0)}$ at $\Delta=0$ versus the frequency difference $W$ between two scattering states in the mirror. (c) Anti-$\mathcal{PT}$ phase transitions in the anti-Bragg atom-dimer cavity. Red-solid and blue-dashed curves correspond to real and imaginary parts of two supermodes. (d) Reflection-dependent atom-dimer cavity. A low-loss cavity supermode is produced by large mirror reflection.}\label{figure2}
\end{figure}

In Fig.~\ref{figure2}(a), we show details of the reflection spectra altered by the atomic coupling $\Omega$. At $\Omega=-\Gamma$ ($W=0$), atoms in the mirror have no coupling~\cite{kannan2023}. Interestingly, degenerate scattering states produce out-of-phase photon components. Due to destructive quantum interference, the anti-Bragg atom-dimer mirror becomes transparent for incident photons with various frequencies, i.e., trivial mirror. Single-peak and two-peak reflection spectra are obtained for $-\Gamma<\Omega\leq 0$ and $\Omega>0$, respectively, enabling the study of reflection-tuned atomic cavities. From Eq.~\eqref{EqR}, we obtain the reflection at $\Delta=0$
\begin{equation}
R^{(0)}= \frac{\Gamma^2 W^2}{\big(\Gamma^2 + \frac{W^2}{4}\big)^2}. \label{EqR0}
\end{equation}
The reflection $R^{(0)}$ responsible for cavity losses~\cite{Chang2012cavity,PhysRevLett.113.163901} is nontrivially changed by $W$. As shown in Fig.~\ref{figure2}(b), $R^{(0)}$ increases with $W$ for single-peak reflection. After reaching unity at $W=2\Gamma$, $R^{(0)}$ is reduced, producing a two-peak reflection spectrum. Therefore, $W$ plays an important role in controlling the reflection spectrum of the atom-dimer mirror.

\textit{Atomic cavity protected by anti-$\mathcal{PT}$ symmetry}.---To explore the relation between atomic mirrors and cavity, we consider two atom-dimer mirrors with Hamiltonian $H_{0} = \sum_{i}\Omega (\sigma_{\mathcal{M}_i,\mathcal{A}_1}^+ \sigma_{\mathcal{M}_i,\mathcal{A}_2}^- + \mathrm{H.c.})$,
where $i=l, r$ denotes the left and right atomic mirror; and $\mathcal{A}_{1}, \mathcal{A}_{2}$ represent mirror atoms. Considering the cavity architecture in Fig.~\ref{cavity}(b), we trace out the degrees of freedom of photons in the waveguide~\cite{PhysRevA.72.063815,PhysRevLett.106.020501,PhysRevA.91.063828,PhysRevA.92.053834,caneva2015quantum,PhysRevA.92.063835,PhysRevA.93.033833}, and obtain an effective non-Hermitian Hamiltonian in the single-excitation subspace $\{|\psi_j\rangle=\sigma_j^+|g_1 g_2 g_3 g_4\rangle\}$ of four mirror atoms~\cite{SupplementalMaterial}
\begin{equation}\label{eqH4}
H_{\mathrm{c}} = (\Omega + \Gamma) s_0 \otimes \tau_x + \Gamma s_x \otimes \tau_0 -i \Gamma s_y \otimes \tau_y -i \Gamma s_0 \otimes \tau_0,
\end{equation}
where $s_n$ and $\tau_n$ ($n=x,y,z$) are Pauli matrices in the space $\{\mathcal{M}_l, \mathcal{M}_r\}$ of two mirrors and the subspace $\{\mathcal{A}_1, \mathcal{A}_2\}$ of mirror atoms, respectively. We have assumed that two atomic mirrors are separated by $\lambda_0$. Equation~\eqref{eqH4} describes the quantum light-mirror interaction in the atomic cavity. To clarify the mechanism of the cavity, we make a unitary transformation to Eq.~\eqref{eqH4}~\cite{SupplementalMaterial},
and simplify the system as two decoupled subsystems $\mathcal{H}_1 \oplus \mathcal{H}_2$ with
\begin{eqnarray}
       \mathcal{H}_1 &=& \begin{pmatrix} \label{EqH1}
                 -\Omega -i \Gamma &  -i \Gamma \\
                  -i \Gamma    & \Omega -i \Gamma
       \end{pmatrix}, \\
       \mathcal{H}_2 &=& \begin{pmatrix}
                 -\Omega-2\Gamma -i \Gamma &  i \Gamma \\
                  i \Gamma    & \Omega+2\Gamma -i \Gamma
       \end{pmatrix}.
\end{eqnarray}
$\mathcal{H}_1$ and $\mathcal{H}_2$ are protected by anti-$\mathcal{PT}$ symmetry $(\mathcal{PT})\mathcal{H}_i(\mathcal{PT})^{-1}=-\mathcal{H}_i$~\cite{PhysRevA.88.053810,PhysRevLett.113.123004,PhysRevA.96.053845,li2019}. Importantly, without using frequency detunings~\cite{peng2016anti,choi2018observation,li2019anti,PhysRevLett.123.193604,PhysRevApplied.13.014053,zhang2020breaking,PhysRevLett.125.147202,PhysRevLett.126.180401,
yang2022radiative}, here anti-$\mathcal{PT}$ phase transitions are produced by atomic couplings, which can uncover novel properties of anti-$\mathcal{PT}$ symmetric systems. Moreover, we avoid the influence of frequency-dependent couplings~\cite{PhysRevLett.120.140404,kannan2020waveguide} on the atom-dimer cavity.

The anti-$\mathcal{PT}$ symmetry inspires us to study the atom-dimer cavity using non-Hermitian theory~\cite{PhysRevLett.80.5243,Ganainy2018non,ozdemir2019parity,ashida2020non}. We diagonalize the system as $H_{\mathrm{c}}=\sum_{n} E_n|\Psi_{n}^R\rangle \langle \Psi_{n}^L|$, with the biorthogonal basis $\langle \Psi_n^L |\Psi_{m}^R \rangle=\delta_{nm}$. The index $j$ labels supermodes of the atomic cavity. In Fig.~\ref{figure2}(c), we show real and imaginary parts of the eigenvalues for two supermodes in the anti-$\mathcal{PT}$-symmetry-protected regimes $W \in (-4\Gamma, 0)\cup (0, 4\Gamma)$. Anti-$\mathcal{PT}$ phase transitions take place at second-order exceptional points $W=0, \pm 4\Gamma$. For $0\leq W\leq 4\Gamma$, the eigenvalues of $\mathcal{H}_1$ correspond to two degenerate supermodes $\Psi_{\pm}$ with decay rates
\begin{equation}\label{EqGammapm}
\Gamma_{\pm}=\Gamma \pm \sqrt{\Gamma^2 - \Omega^2}.
\end{equation}
At $\Omega=0$, a long-living supermode exists in the atom-dimer cavity with mirrors having complete reflection $R^{(0)}$ at the central frequency. This supermode plays the role of a cavity mode~\cite{mirhosseini2019cavity}, and becomes dissipative for reduced $R^{(0)}$. By solving Eq.~(\ref{EqR0}) for $\Omega$ and substituting the solutions to Eq.~(\ref{EqGammapm}), we obtain reflection-tuned energy levels and decay rates of the supermodes $\Psi_{\pm}$ for $W\geq 2\Gamma$, as shown in Fig.~\ref{figure2}(d). For weak reflection, these two supermodes have equal decay rate $\Gamma$, same as individual mirror atoms. However, large reflection leads to an anti-$\mathcal{PT}$ phase transition, producing cavity supermodes with controlled loss.

\begin{figure}[b]
\includegraphics[width=8.5cm]{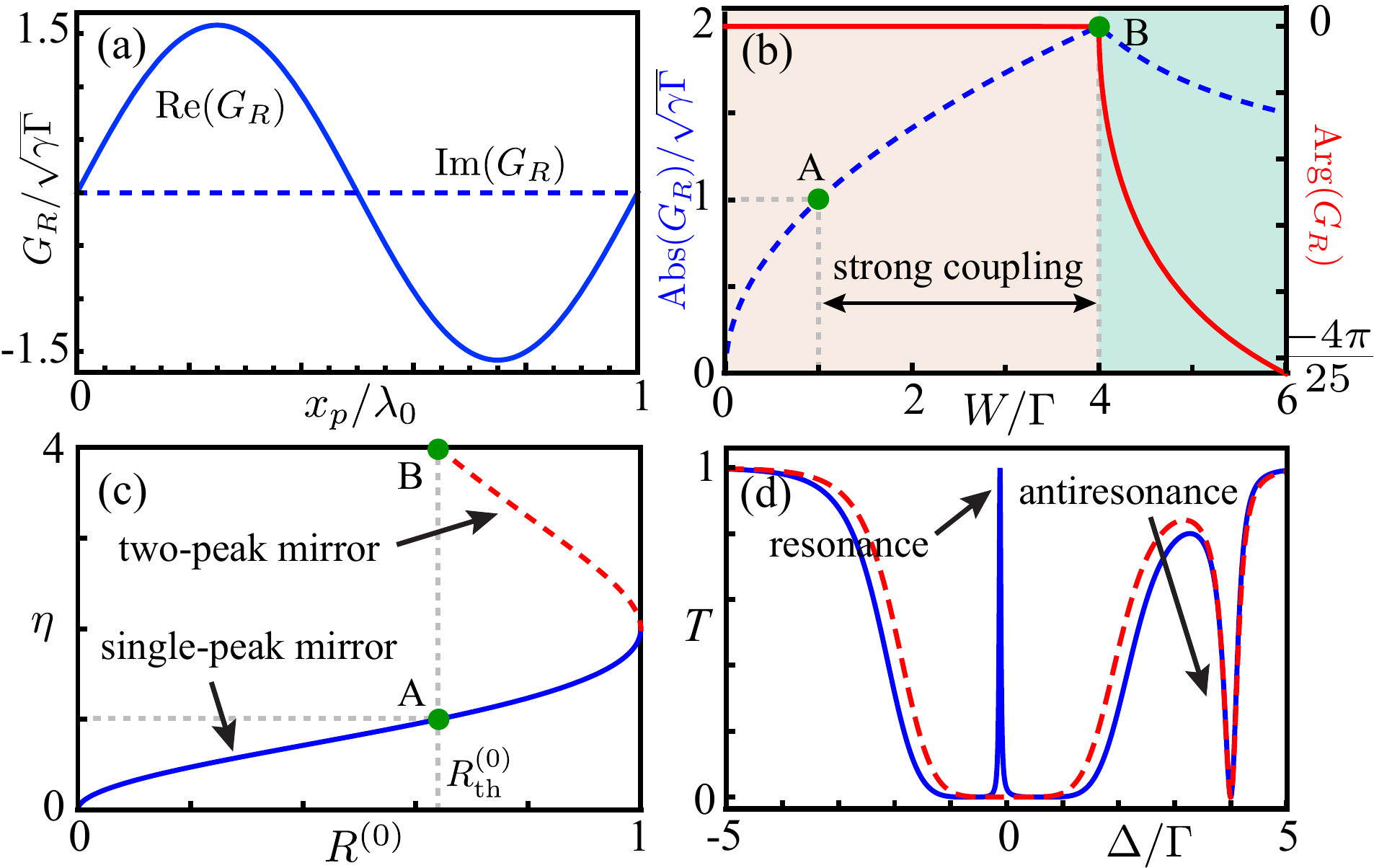}
\caption{(a) Coupling between probe atom and the slow-decay supermode. The solid curve and dashed line denote the real and imaginary parts of the coupling $G_R$. The horizontal axis represents the position of the probe atom in the cavity. (b) Absolute value and argument of cavity-atom coupling $G_R$ for the slow-decay supermode. The probe atom is located at $x_p=0.25\lambda_0$. (c) Atomic cavity field changed by the mirror reflection. (d) Transmission spectrum of the far-detuned probe atom. Red-dashed and blue-solid curves correspond to $\Omega=0$ and $0.2\Gamma$, respectively.}\label{figure3}
\end{figure}

\textit{Reflection threshold for strong cavity-atom coupling}.---Besides photon loss, the mode profile is a fundamental property of cavities~\cite{PhysRevA.64.033804,PhysRevE.65.016608}. In optical cavities, the cavity-atom coupling is proportional to the cavity mode electric field, which is related to boundary conditions imposed by mirrors~\cite{Friedler2009solid,PhysRevB.98.121306}. To study cavity fields affected by atomic mirrors, we consider a probe atom [see Fig.~\ref{cavity}(a)]. The atom-cavity interaction is described by $H_{\mathrm{int}}=(\delta \omega -i \gamma)\sigma_p^+\sigma_p^- -i\sqrt{\gamma \Gamma} \sum_{j=1}^4 e^{i\phi_{j}} (\sigma_{j}^+ \sigma_{p}^- + \sigma_{p}^+ \sigma_{j}^-)$. Here, $\delta\omega = \omega_p -\omega_0$ is the detuning between the probe atom and mirror atoms; $\gamma$ is decay rate of the probe atom; $\sigma_j^{\pm}$ are operators of the $j$th mirror atoms; and $\phi_{j}=2\pi |x_j-x_p|/\lambda_0$~\cite{albrecht2019subradiant,PhysRevLett.122.203605,PhysRevLett.123.253601}. The probe atom is placed at $x_p=\lambda_0/4$ in the cavity, such that the couplings vanish between the probe atom and supermodes unprotected by anti-$\mathcal{PT}$ symmetry~\cite{SupplementalMaterial}.  By writing the whole Hamiltonian $\tilde{H}=H_{\mathrm{c}} + H_{\mathrm{int}}$ in terms of two supermodes $\Psi_{\pm}$, we obtain
\begin{eqnarray}
       \tilde{H} = \begin{pmatrix}
                      -i \Gamma_{-} & 0 & G_L \\
                       0     & -i \Gamma_{+} & V_L \\
                      G_R     &  V_R  & \delta \omega -i \gamma
       \end{pmatrix}, \label{Eqeff}
\end{eqnarray}
where the couplings are $G_L = -i\sqrt{\gamma \Gamma}\sum_j e^{i \phi_j} \langle \Psi_{-}^L|\psi_j\rangle$,
$G_R = -i\sqrt{\gamma \Gamma}\sum_j e^{i \phi_j} \langle \psi_{j}|\Psi_-^R\rangle$, and similarly for $V_{L,R}$ by considering the supermode $\Psi_{+}$. With the protection of anti-$\mathcal{PT}$ symmetry, the probe atom is coherently and dissipatively coupled to the slow- and fast-decay supermodes with the same coupling strength, i.e., $\mathrm{Im}[G_R]=0$ and $V_R=i G_R$. As shown in Fig.~\ref{figure3}(a), the probe atom is maximally coupled to the supermode $\Psi_{-}$ at $x_p=\lambda_0/4$. We find $G_L=G_R/\sqrt{1-\Omega^2/\Gamma^2}$ and $V_L=V_R/\sqrt{1-\Omega^2/\Gamma^2}$, which diverge at exceptional points. Hence, $G_{R}$ and $V_{R}$ characterize the effective fields of the atomic cavity.

Figure~\ref{figure3}(b) presents the absolute value and argument of the coupling $G_R$ between the atom and the slow-decay supermode. Indeed, the coupling is coherent for $0 \leq W \leq 4\Gamma$, and its concise form is
\begin{equation} \label{Eqcoupling}
G_R=\sqrt{\gamma W}.
\end{equation}
This equation uncovers an intrinsic relation between mirror reflection and atom-dimer cavity: the frequency difference $W$ between mirror's scattering states determines the atom-cavity coupling. At point $A$, $W=\Gamma$ indicates the emergence of enhanced cavity-atom coupling with respect to the couplings between probe atom and mirror atoms. At point $B$ with $W=4\Gamma$, the cavity undergoes a phase transition and the coherent cavity-atom coupling reaches its maximum. Therefore, the parameter space between $A$ and $B$ is the strong-coupling regime.

To gain further insight into the relation between mirror reflection and the atom-dimer cavity, we define a coupling factor $\eta=G_R^2/\gamma \Gamma$ in terms of $R^{(0)}$ via Eq.~(\ref{EqR0}). The coupling factor corresponding to atomic mirrors with single-peak reflection spectra is
$\eta=2(1-\sqrt{1-R^{(0)}})/\sqrt{R^{(0)}}$. As shown in Fig.~\ref{figure3}(c), $\eta$ monotonically increases with $R^{(0)}$, showing a reflection-controlled cavity field~\cite{PhysRevB.87.115419}. For two-peak mirrors, the coupling factor becomes $\eta=2\sqrt{R^{(0)}}/(1-\sqrt{1-R^{(0)}})$. It increases with growing $W$ even though $R^{(0)}$ declines. However, when $R^{(0)}$ reduces to a critical value at point $B$, the coupling $G_R$ becomes dissipative. Interestingly, we find that points $A$ and $B$ correspond to a reflection threshold $R_{\mathrm{th}}^{(0)} = 0.64$. \emph{The strong coherent coupling requires $R^{(0)} > R_{\mathrm{th}}^{(0)}$}.

In Fig.~\ref{figure3}(d), we show transmission spectrum of the detuned cavity-atom system. The antiresonance in the transmission is due to the probe atom~\cite{PhysRevLett.95.213001,chang2007single,PhysRevLett.101.100501}. Owing to the anti-Bragg scattering, dissipative supermodes in the atomic cavity produce a transmissionless spectrum for $\Omega=0$. Weak atomic coupling leads to a photon transmission amplitude $t \approx \Gamma_{-}/[i(\Delta+\Delta_p) + \Gamma_{-}]$, where $\Delta_p$ is the frequency shift induced by the probe atom. In contrast to conventional subradiant states that inhibit photon transmission~\cite{nie2020nonreciprocal}, the slow-decay supermode $\Psi_{-}$ enhances photon transmission. This cavity-like behavior~\cite{PhysRevLett.119.093601} makes it useful to detect atomic cavity QED via photon transport in a waveguide.

\begin{figure}[t]
\includegraphics[width=8.5cm]{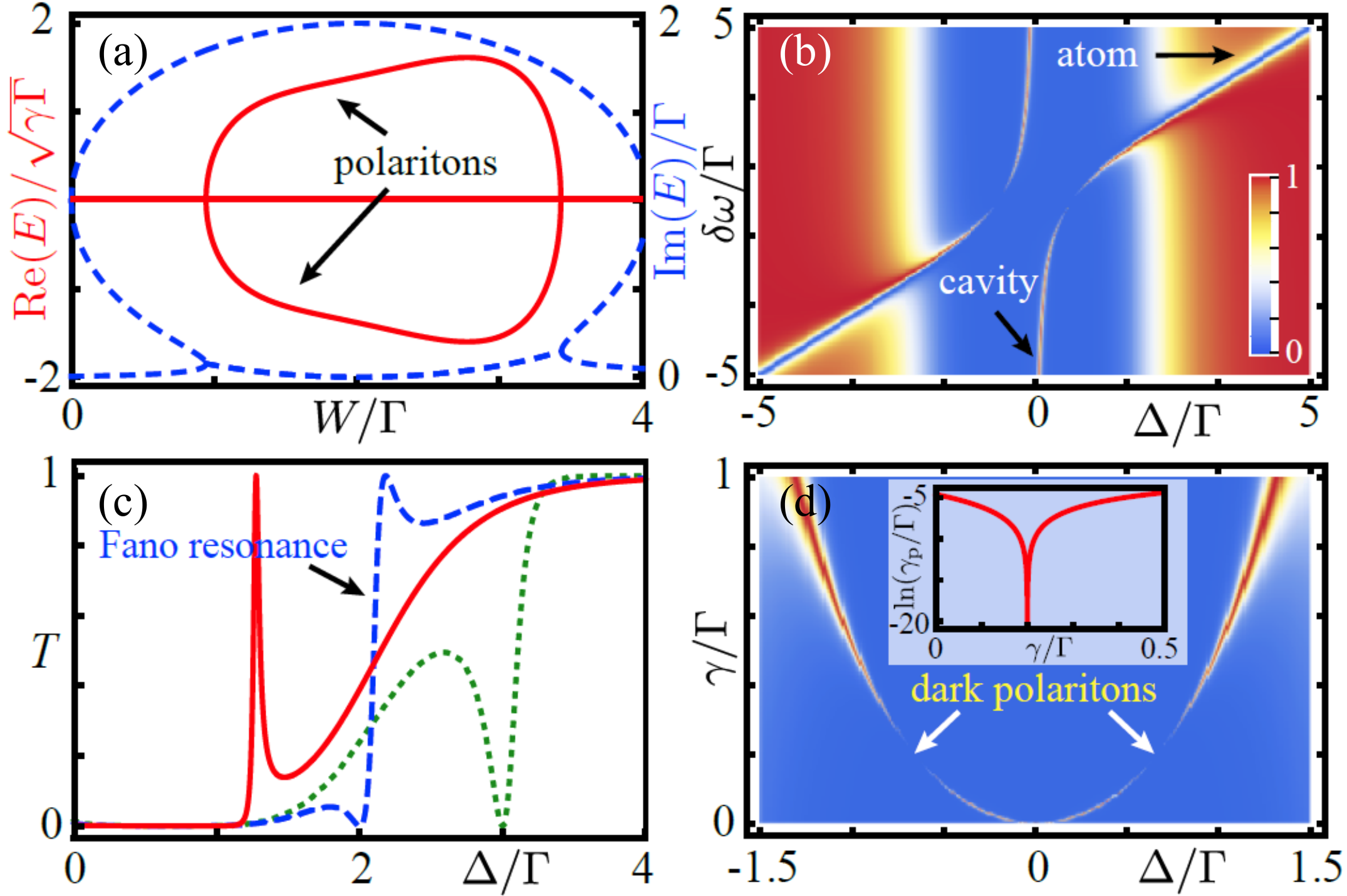}
\caption{(a) Cavity-atom polaritons are formed with $\gamma=0.005\Gamma$. (b) Transmission detection of polaritons for $\gamma=0.2\Gamma$. (c) Polariton-induced Fano resonance. Red-solid, blue-dashed and green-dotted curves correspond to $\delta\omega/\Gamma=1, 2$ and $3$, respectively. (d) $\gamma$-dependent polaritons. The inset shows $\ln(\gamma_{p}/\Gamma)$ versus $\gamma$, where $\gamma_p$ denotes decay rate of polaritons. We consider $\Omega=0.2\Gamma$ in (b-d), $\gamma=0.2\Gamma$ in (b,c) and $\delta\omega=0$ in (a,d).}\label{figure4}
\end{figure}

\textit{Mirror-controlled cavity-atom polaritons}.---In cavity QED, polaritons can be produced with interacting photons and atoms/excitons~\cite{Basov2021,ghosh2022microcavity}. In our system, non-Hermitian cavity-atom interactions in Eq.~(\ref{Eqeff}) control the formation of cavity-atom polaritons. Solutions of the equation $\mathrm{det}(\tilde{H}-E) =0$ can be derived using Cardano's formula~\cite{jing2017high,SupplementalMaterial}. In Fig.~\ref{figure4}(a), we show the eigenspectrum of the cavity-atom system. In the strong-coupling regime, the probe atom hybridizes intensely with the slow-decay supermode, giving rise to two equally decaying polaritons. We find that at the condition $\Omega=\gamma$, the eigenvalues are $E_{1,2} = \pm \sqrt{\Omega^2 + 2\Gamma \gamma}$ and $E_3 = -i(2\Gamma + \gamma)$. Here, $E_{1,2}$ correspond to two polaritons without dissipation, i.e., dark polaritons; while $E_3$ contains the whole dissipation. Therefore, the fast-decay supermode is crucial for tuning decay rates of polaritons. Moreover, the dark polaritons are only generated in cavities with two-peak atom-dimer mirrors, because $\gamma>0$. As shown in Fig.~\ref{figure4}(b), the cavity-atom polaritons can be detected by cavity transmission. The vanishing signals at resonance represent dark polaritons. The dark polaritons studied here emerge from the novel non-Hermitian interaction between the probe atom and the two-mode atomic cavity, and cannot be produced by single-mode optical cavities~\cite{hennessy2007quantum}.

In Fig.~\ref{figure4}(c), we show the transmission of one polariton for various atom-cavity frequency detunings. A Fano resonance is found around $\delta\omega=2\Gamma$. Considering distinct spectroscopic signatures of the probe atom and the cavity supermode $\Psi_{-}$ shown in Fig.~\ref{figure3}(d), the Fano resonance reveals the half-matter half-light nature of polaritons, useful for optical switching and sensing~\cite{limonov2017fano}. Figure~\ref{figure4}(d) displays the transmission spectrum of $\gamma$-dependent polaritons. The frequencies of polaritons agree with Eq.~(\ref{Eqcoupling}). This confirms the effective cavity-atom coupling represented by $G_R$. The inset shows the linewidth of polaritons versus $\gamma$ with a minimum at $\gamma=\Omega$.

\begin{figure}[t]
\includegraphics[width=8.5cm]{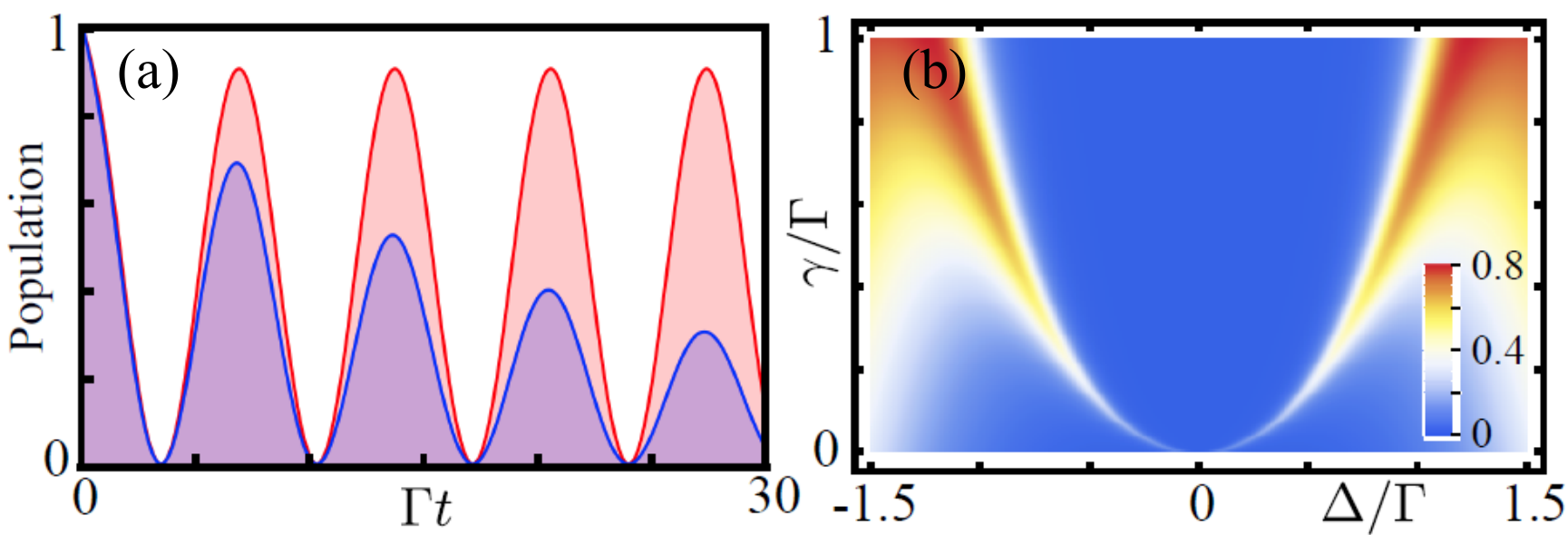}
\caption{(a) Rabi oscillations of a probe atom due to the dark polaritons. We consider an atomic coupling $\Omega=0.1\Gamma$ and decay rate $\gamma=0.1\Gamma$ of probe atom. Red and blue curves respond to free-space loss $\gamma'=0$ and $\gamma'=0.02\Gamma$, respectively. (b) Transmission spectrum of cavity-atom polaritons for $\Omega=-0.2\Gamma$ and free-space loss $\gamma'=0.02\Gamma$.}\label{figure5}
\end{figure}

\textit{Applications of the atom-dimer cavity}.---Slow-decay states, or subradiant states, are useful for quantum information storage~\cite{PhysRevLett.115.243602,PhysRevX.7.031024}. However, it is challenging to access and manipulate these subtle many-body states~\cite{PhysRevX.11.021031,PhysRevResearch.4.013110,PhysRevLett.129.253601}. The atom-dimer cavity provides an interface between a single atom and a subradiant state. We show the persistent Rabi oscillations of the probe atom in Fig.~\ref{figure5}(a), produced by dark polaritons. After considering the free-space loss~\cite{mirhosseini2019cavity}, the population transfer is still efficient. Due to this cavity-atom interface, quantum information can be flexibly stored in and retrieved from the atom-dimer cavity. Figure~\ref{figure5}(b) shows the transmission spectrum of polaritons for a cavity with single-peak mirrors. Different from the two-peak atomic cavity, only bright polaritons are formed.

\textit{Conclusions}.---In this work, we study an open cavity with tunable atom-dimer mirrors. Atomic couplings in mirrors nontrivially control the anti-$\mathcal{PT}$-symmetry protected cavity and produce two reflection-dependent degenerate supermodes. The coherent coupling between the slow-decay supermode and the probe atom is related to frequency difference between mirror's scattering states. We propose a non-Hermitian cavity QED theory, and identify a reflection threshold for strong cavity-atom coupling. The roles played by the slow- and fast-decay supermodes are clarified for realizing dark polaritons. Our work presents a novel cavity in a coupling-controlled anti-$\mathcal{PT}$ symmetric system, which allows to study non-Hermitian light-matter interaction.

\begin{acknowledgments}
The authors thank Stephen Hughes and Zongping Gong for critical readings. W.N. is supported by the National Natural Science Foundation of China (Grant No. 12105025). T.S. acknowledges support from National Key Research and Development Program of China (Grant No. 2017YFA0718304) and the NSFC (Grants No. 11974363, No. 12135018, and No. 12047503). Y.X.L. is supported by the NSFC under Grant No. 11874037, the Key-Area Research and Development Program of GuangDong Province under Grant No. 2018B030326001, and the Innovation Program for Quantum Science and Technology (Grant No. 2021ZD0300200). F.N. is supported in part by: Nippon Telegraph and Telephone Corporation (NTT) Research, the Japan Science and Technology Agency (JST) [via the Quantum Leap Flagship Program (Q-LEAP), and the Moonshot R\&D Grant Number JPMJMS2061], the Asian Office of Aerospace Research and Development (AOARD) (via Grant No. FA2386-20-1-4069), and the Foundational Questions Institute Fund (FQXi) via Grant No. FQXi-IAF19-06.
\end{acknowledgments}

\begin{thebibliography}{99}%
\makeatletter
\providecommand \@ifxundefined [1]{%
 \@ifx{#1\undefined}
}%
\providecommand \@ifnum [1]{%
 \ifnum #1\expandafter \@firstoftwo
 \else \expandafter \@secondoftwo
 \fi
}%
\providecommand \@ifx [1]{%
 \ifx #1\expandafter \@firstoftwo
 \else \expandafter \@secondoftwo
 \fi
}%
\providecommand \natexlab [1]{#1}%
\providecommand \enquote  [1]{``#1''}%
\providecommand \bibnamefont  [1]{#1}%
\providecommand \bibfnamefont [1]{#1}%
\providecommand \citenamefont [1]{#1}%
\providecommand \href@noop [0]{\@secondoftwo}%
\providecommand \href [0]{\begingroup \@sanitize@url \@href}%
\providecommand \@href[1]{\@@startlink{#1}\@@href}%
\providecommand \@@href[1]{\endgroup#1\@@endlink}%
\providecommand \@sanitize@url [0]{\catcode `\\12\catcode `\$12\catcode
  `\&12\catcode `\#12\catcode `\^12\catcode `\_12\catcode `\%12\relax}%
\providecommand \@@startlink[1]{}%
\providecommand \@@endlink[0]{}%
\providecommand \url  [0]{\begingroup\@sanitize@url \@url }%
\providecommand \@url [1]{\endgroup\@href {#1}{\urlprefix }}%
\providecommand \urlprefix  [0]{URL }%
\providecommand \Eprint [0]{\href }%
\providecommand \doibase [0]{http://dx.doi.org/}%
\providecommand \selectlanguage [0]{\@gobble}%
\providecommand \bibinfo  [0]{\@secondoftwo}%
\providecommand \bibfield  [0]{\@secondoftwo}%
\providecommand \translation [1]{[#1]}%
\providecommand \BibitemOpen [0]{}%
\providecommand \bibitemStop [0]{}%
\providecommand \bibitemNoStop [0]{.\EOS\space}%
\providecommand \EOS [0]{\spacefactor3000\relax}%
\providecommand \BibitemShut  [1]{\csname bibitem#1\endcsname}%
\let\auto@bib@innerbib\@empty
\bibitem [{\citenamefont {Scully}\ and\ \citenamefont
  {Zubairy}(1997)}]{scully1997quantum}%
  \BibitemOpen
  \bibfield  {author} {\bibinfo {author} {\bibfnamefont {M.~O.}\ \bibnamefont
  {Scully}}\ and\ \bibinfo {author} {\bibfnamefont {M.~S.}\ \bibnamefont
  {Zubairy}},\ }\href {\doibase 10.1017/CBO9780511813993} {\emph {\bibinfo
  {title} {Quantum Optics}}}\ (\bibinfo  {publisher} {Cambridge University
  Press, Cambridge, England},\ \bibinfo {year} {1997})\BibitemShut {NoStop}%
\bibitem [{\citenamefont {Agarwal}(2012)}]{Agarwal2012}%
  \BibitemOpen
  \bibfield  {author} {\bibinfo {author} {\bibfnamefont {G.~S.}\ \bibnamefont
  {Agarwal}},\ }\href {\doibase 10.1017/CBO9781139035170} {\emph {\bibinfo
  {title} {Quantum Optics}}}\ (\bibinfo  {publisher} {Cambridge University
  Press, Cambridge, England},\ \bibinfo {year} {2012})\BibitemShut {NoStop}%
\bibitem [{\citenamefont {Haroche}\ and\ \citenamefont
  {Raimond}(2006)}]{Haroche2006book}%
  \BibitemOpen
  \bibfield  {author} {\bibinfo {author} {\bibfnamefont {S.}~\bibnamefont
  {Haroche}}\ and\ \bibinfo {author} {\bibfnamefont {J.-M.}\ \bibnamefont
  {Raimond}},\ }\href {\doibase 10.1093/acprof:oso/9780198509141.001.0001}
  {\emph {\bibinfo {title} {Exploring the Quantum: Atoms, Cavities, and
  Photons}}}\ (\bibinfo  {publisher} {Oxford University Press, New York},\
  \bibinfo {year} {2006})\BibitemShut {NoStop}%
\bibitem [{\citenamefont {Carusotto}\ and\ \citenamefont
  {Ciuti}(2013)}]{RevModPhys.85.299}%
  \BibitemOpen
  \bibfield  {author} {\bibinfo {author} {\bibfnamefont {I.}~\bibnamefont
  {Carusotto}}\ and\ \bibinfo {author} {\bibfnamefont {C.}~\bibnamefont
  {Ciuti}},\ }\bibfield  {title} {\emph {\bibinfo {title} {{Quantum fluids of
  light}},\ }}\href {\doibase 10.1103/RevModPhys.85.299} {\bibfield  {journal}
  {\bibinfo  {journal} {Rev. Mod. Phys.}\ }\textbf {\bibinfo {volume} {85}},\
  \bibinfo {pages} {299} (\bibinfo {year} {2013})}\BibitemShut {NoStop}%
\bibitem [{\citenamefont {Kavokin}\ \emph {et~al.}(2017)\citenamefont
  {Kavokin}, \citenamefont {Baumberg}, \citenamefont {Malpuech},\ and\
  \citenamefont {Laussy}}]{kavokin2017}%
  \BibitemOpen
  \bibfield  {author} {\bibinfo {author} {\bibfnamefont {A.~V.}\ \bibnamefont
  {Kavokin}}, \bibinfo {author} {\bibfnamefont {J.~J.}\ \bibnamefont
  {Baumberg}}, \bibinfo {author} {\bibfnamefont {G.}~\bibnamefont {Malpuech}},
  \ and\ \bibinfo {author} {\bibfnamefont {F.~P.}\ \bibnamefont {Laussy}},\
  }\href {\doibase 10.1093/oso/9780198782995.001.0001} {\emph {\bibinfo {title}
  {Microcavities}}}\ (\bibinfo  {publisher} {Oxford University Press},\
  \bibinfo {year} {2017})\BibitemShut {NoStop}%
\bibitem [{\citenamefont {Zhou}\ \emph
  {et~al.}(2008{\natexlab{a}})\citenamefont {Zhou}, \citenamefont {Dong},
  \citenamefont {Liu}, \citenamefont {Sun},\ and\ \citenamefont
  {Nori}}]{PhysRevA.78.063827}%
  \BibitemOpen
  \bibfield  {author} {\bibinfo {author} {\bibfnamefont {L.}~\bibnamefont
  {Zhou}}, \bibinfo {author} {\bibfnamefont {H.}~\bibnamefont {Dong}}, \bibinfo
  {author} {\bibfnamefont {Y.-X.}\ \bibnamefont {Liu}}, \bibinfo {author}
  {\bibfnamefont {C.~P.}\ \bibnamefont {Sun}}, \ and\ \bibinfo {author}
  {\bibfnamefont {F.}~\bibnamefont {Nori}},\ }\bibfield  {title} {\emph
  {\bibinfo {title} {{Quantum supercavity with atomic mirrors}},\ }}\href
  {\doibase 10.1103/PhysRevA.78.063827} {\bibfield  {journal} {\bibinfo
  {journal} {Phys. Rev. A}\ }\textbf {\bibinfo {volume} {78}},\ \bibinfo
  {pages} {063827} (\bibinfo {year} {2008}{\natexlab{a}})}\BibitemShut
  {NoStop}%
\bibitem [{\citenamefont {Chang}\ \emph {et~al.}(2011)\citenamefont {Chang},
  \citenamefont {Gong},\ and\ \citenamefont {Sun}}]{PhysRevA.83.013825}%
  \BibitemOpen
  \bibfield  {author} {\bibinfo {author} {\bibfnamefont {Y.}~\bibnamefont
  {Chang}}, \bibinfo {author} {\bibfnamefont {Z.~R.}\ \bibnamefont {Gong}}, \
  and\ \bibinfo {author} {\bibfnamefont {C.~P.}\ \bibnamefont {Sun}},\
  }\bibfield  {title} {\emph {\bibinfo {title} {{Multiatomic mirror for perfect
  reflection of single photons in a wide band of frequency}},\ }}\href
  {\doibase 10.1103/PhysRevA.83.013825} {\bibfield  {journal} {\bibinfo
  {journal} {Phys. Rev. A}\ }\textbf {\bibinfo {volume} {83}},\ \bibinfo
  {pages} {013825} (\bibinfo {year} {2011})}\BibitemShut {NoStop}%
\bibitem [{\citenamefont {H\'etet}\ \emph {et~al.}(2011)\citenamefont
  {H\'etet}, \citenamefont {Slodi\ifmmode~\check{c}\else \v{c}\fi{}ka},
  \citenamefont {Hennrich},\ and\ \citenamefont
  {Blatt}}]{PhysRevLett.107.133002}%
  \BibitemOpen
  \bibfield  {author} {\bibinfo {author} {\bibfnamefont {G.}~\bibnamefont
  {H\'etet}}, \bibinfo {author} {\bibfnamefont {L.}~\bibnamefont
  {Slodi\ifmmode~\check{c}\else \v{c}\fi{}ka}}, \bibinfo {author}
  {\bibfnamefont {M.}~\bibnamefont {Hennrich}}, \ and\ \bibinfo {author}
  {\bibfnamefont {R.}~\bibnamefont {Blatt}},\ }\bibfield  {title} {\emph
  {\bibinfo {title} {{Single Atom as a Mirror of an Optical Cavity}},\ }}\href
  {\doibase 10.1103/PhysRevLett.107.133002} {\bibfield  {journal} {\bibinfo
  {journal} {Phys. Rev. Lett.}\ }\textbf {\bibinfo {volume} {107}},\ \bibinfo
  {pages} {133002} (\bibinfo {year} {2011})}\BibitemShut {NoStop}%
\bibitem [{\citenamefont {Chang}\ \emph {et~al.}(2012)\citenamefont {Chang},
  \citenamefont {Jiang}, \citenamefont {Gorshkov},\ and\ \citenamefont
  {Kimble}}]{Chang2012cavity}%
  \BibitemOpen
  \bibfield  {author} {\bibinfo {author} {\bibfnamefont {D.~E.}\ \bibnamefont
  {Chang}}, \bibinfo {author} {\bibfnamefont {L.}~\bibnamefont {Jiang}},
  \bibinfo {author} {\bibfnamefont {A.}~\bibnamefont {Gorshkov}}, \ and\
  \bibinfo {author} {\bibfnamefont {H.}~\bibnamefont {Kimble}},\ }\bibfield
  {title} {\emph {\bibinfo {title} {{Cavity QED with atomic mirrors}},\ }}\href
  {\doibase 10.1088/1367-2630/14/6/063003} {\bibfield  {journal} {\bibinfo
  {journal} {New J. Phys.}\ }\textbf {\bibinfo {volume} {14}},\ \bibinfo
  {pages} {063003} (\bibinfo {year} {2012})}\BibitemShut {NoStop}%
\bibitem [{\citenamefont {Fratini}\ \emph {et~al.}(2014)\citenamefont
  {Fratini}, \citenamefont {Mascarenhas}, \citenamefont {Safari}, \citenamefont
  {Poizat}, \citenamefont {Valente}, \citenamefont {Auff\`eves}, \citenamefont
  {Gerace},\ and\ \citenamefont {Santos}}]{PhysRevLett.113.243601}%
  \BibitemOpen
  \bibfield  {author} {\bibinfo {author} {\bibfnamefont {F.}~\bibnamefont
  {Fratini}}, \bibinfo {author} {\bibfnamefont {E.}~\bibnamefont
  {Mascarenhas}}, \bibinfo {author} {\bibfnamefont {L.}~\bibnamefont {Safari}},
  \bibinfo {author} {\bibfnamefont {J.-P.}\ \bibnamefont {Poizat}}, \bibinfo
  {author} {\bibfnamefont {D.}~\bibnamefont {Valente}}, \bibinfo {author}
  {\bibfnamefont {A.}~\bibnamefont {Auff\`eves}}, \bibinfo {author}
  {\bibfnamefont {D.}~\bibnamefont {Gerace}}, \ and\ \bibinfo {author}
  {\bibfnamefont {M.~F.}\ \bibnamefont {Santos}},\ }\bibfield  {title} {\emph
  {\bibinfo {title} {{Fabry-Perot Interferometer with Quantum Mirrors:
  Nonlinear Light Transport and Rectification}},\ }}\href {\doibase
  10.1103/PhysRevLett.113.243601} {\bibfield  {journal} {\bibinfo  {journal}
  {Phys. Rev. Lett.}\ }\textbf {\bibinfo {volume} {113}},\ \bibinfo {pages}
  {243601} (\bibinfo {year} {2014})}\BibitemShut {NoStop}%
\bibitem [{\citenamefont {Hoi}\ \emph {et~al.}(2015)\citenamefont {Hoi},
  \citenamefont {Kockum}, \citenamefont {Tornberg}, \citenamefont
  {Pourkabirian}, \citenamefont {Johansson}, \citenamefont {Delsing},\ and\
  \citenamefont {Wilson}}]{hoi2015probing}%
  \BibitemOpen
  \bibfield  {author} {\bibinfo {author} {\bibfnamefont {I.-C.}\ \bibnamefont
  {Hoi}}, \bibinfo {author} {\bibfnamefont {A.~F.}\ \bibnamefont {Kockum}},
  \bibinfo {author} {\bibfnamefont {L.}~\bibnamefont {Tornberg}}, \bibinfo
  {author} {\bibfnamefont {A.}~\bibnamefont {Pourkabirian}}, \bibinfo {author}
  {\bibfnamefont {G.}~\bibnamefont {Johansson}}, \bibinfo {author}
  {\bibfnamefont {P.}~\bibnamefont {Delsing}}, \ and\ \bibinfo {author}
  {\bibfnamefont {C. M.}~\bibnamefont {Wilson}},\ }\bibfield  {title} {\emph
  {\bibinfo {title} {{Probing the quantum vacuum with an artificial atom in
  front of a mirror}},\ }}\href {\doibase 10.1038/nphys3484} {\bibfield
  {journal} {\bibinfo  {journal} {Nat. Phys.}\ }\textbf {\bibinfo {volume}
  {11}},\ \bibinfo {pages} {1045} (\bibinfo {year} {2015})}\BibitemShut
  {NoStop}%
\bibitem [{\citenamefont {Wen}\ \emph {et~al.}(2019)\citenamefont {Wen},
  \citenamefont {Lin}, \citenamefont {Kockum}, \citenamefont {Suri},
  \citenamefont {Ian}, \citenamefont {Chen}, \citenamefont {Mao}, \citenamefont
  {Chiu}, \citenamefont {Delsing}, \citenamefont {Nori}, \citenamefont {Lin},\
  and\ \citenamefont {Hoi}}]{PhysRevLett.123.233602}%
  \BibitemOpen
  \bibfield  {author} {\bibinfo {author} {\bibfnamefont {P.~Y.}\ \bibnamefont
  {Wen}}, \bibinfo {author} {\bibfnamefont {K.-T.}\ \bibnamefont {Lin}},
  \bibinfo {author} {\bibfnamefont {A.~F.}\ \bibnamefont {Kockum}}, \bibinfo
  {author} {\bibfnamefont {B.}~\bibnamefont {Suri}}, \bibinfo {author}
  {\bibfnamefont {H.}~\bibnamefont {Ian}}, \bibinfo {author} {\bibfnamefont
  {J.~C.}\ \bibnamefont {Chen}}, \bibinfo {author} {\bibfnamefont {S.~Y.}\
  \bibnamefont {Mao}}, \bibinfo {author} {\bibfnamefont {C.~C.}\ \bibnamefont
  {Chiu}}, \bibinfo {author} {\bibfnamefont {P.}~\bibnamefont {Delsing}},
  \bibinfo {author} {\bibfnamefont {F.}~\bibnamefont {Nori}}, \bibinfo {author}
  {\bibfnamefont {G.-D.}\ \bibnamefont {Lin}}, \ and\ \bibinfo {author}
  {\bibfnamefont {I.-C.}\ \bibnamefont {Hoi}},\ }\bibfield  {title} {\emph
  {\bibinfo {title} {{Large Collective Lamb Shift of Two Distant
  Superconducting Artificial Atoms}},\ }}\href {\doibase
  10.1103/PhysRevLett.123.233602} {\bibfield  {journal} {\bibinfo  {journal}
  {Phys. Rev. Lett.}\ }\textbf {\bibinfo {volume} {123}},\ \bibinfo {pages}
  {233602} (\bibinfo {year} {2019})}\BibitemShut {NoStop}%
\bibitem [{\citenamefont {Domokos}\ \emph {et~al.}(2002)\citenamefont
  {Domokos}, \citenamefont {Horak},\ and\ \citenamefont
  {Ritsch}}]{PhysRevA.65.033832}%
  \BibitemOpen
  \bibfield  {author} {\bibinfo {author} {\bibfnamefont {P.}~\bibnamefont
  {Domokos}}, \bibinfo {author} {\bibfnamefont {P.}~\bibnamefont {Horak}}, \
  and\ \bibinfo {author} {\bibfnamefont {H.}~\bibnamefont {Ritsch}},\
  }\bibfield  {title} {\emph {\bibinfo {title} {{Quantum description of
  light-pulse scattering on a single atom in waveguides}},\ }}\href {\doibase
  10.1103/PhysRevA.65.033832} {\bibfield  {journal} {\bibinfo  {journal} {Phys.
  Rev. A}\ }\textbf {\bibinfo {volume} {65}},\ \bibinfo {pages} {033832}
  (\bibinfo {year} {2002})}\BibitemShut {NoStop}%
\bibitem [{\citenamefont {Shen}\ and\ \citenamefont
  {Fan}(2005)}]{PhysRevLett.95.213001}%
  \BibitemOpen
  \bibfield  {author} {\bibinfo {author} {\bibfnamefont {J.-T.}\ \bibnamefont
  {Shen}}\ and\ \bibinfo {author} {\bibfnamefont {S.}~\bibnamefont {Fan}},\
  }\bibfield  {title} {\emph {\bibinfo {title} {{Coherent Single Photon
  Transport in a One-Dimensional Waveguide Coupled with Superconducting Quantum
  Bits}},\ }}\href {\doibase 10.1103/PhysRevLett.95.213001} {\bibfield
  {journal} {\bibinfo  {journal} {Phys. Rev. Lett.}\ }\textbf {\bibinfo
  {volume} {95}},\ \bibinfo {pages} {213001} (\bibinfo {year}
  {2005})}\BibitemShut {NoStop}%
\bibitem [{\citenamefont {Chang}\ \emph {et~al.}(2007)\citenamefont {Chang},
  \citenamefont {S{\o}rensen}, \citenamefont {Demler},\ and\ \citenamefont
  {Lukin}}]{chang2007single}%
  \BibitemOpen
  \bibfield  {author} {\bibinfo {author} {\bibfnamefont {D.~E.}\ \bibnamefont
  {Chang}}, \bibinfo {author} {\bibfnamefont {A.~S.}\ \bibnamefont
  {S{\o}rensen}}, \bibinfo {author} {\bibfnamefont {E.~A.}\ \bibnamefont
  {Demler}}, \ and\ \bibinfo {author} {\bibfnamefont {M.~D.}\ \bibnamefont
  {Lukin}},\ }\bibfield  {title} {\emph {\bibinfo {title} {{A single-photon
  transistor using nanoscale surface plasmons}},\ }}\href {\doibase
  10.1038/nphys708} {\bibfield  {journal} {\bibinfo  {journal} {Nat. Phys.}\
  }\textbf {\bibinfo {volume} {3}},\ \bibinfo {pages} {807} (\bibinfo {year}
  {2007})}\BibitemShut {NoStop}%
\bibitem [{\citenamefont {Zhou}\ \emph
  {et~al.}(2008{\natexlab{b}})\citenamefont {Zhou}, \citenamefont {Gong},
  \citenamefont {Liu}, \citenamefont {Sun},\ and\ \citenamefont
  {Nori}}]{PhysRevLett.101.100501}%
  \BibitemOpen
  \bibfield  {author} {\bibinfo {author} {\bibfnamefont {L.}~\bibnamefont
  {Zhou}}, \bibinfo {author} {\bibfnamefont {Z.~R.}\ \bibnamefont {Gong}},
  \bibinfo {author} {\bibfnamefont {Y.-X.}\ \bibnamefont {Liu}}, \bibinfo
  {author} {\bibfnamefont {C.~P.}\ \bibnamefont {Sun}}, \ and\ \bibinfo
  {author} {\bibfnamefont {F.}~\bibnamefont {Nori}},\ }\bibfield  {title}
  {\emph {\bibinfo {title} {{Controllable Scattering of a Single Photon inside
  a One-Dimensional Resonator Waveguide}},\ }}\href {\doibase
  10.1103/PhysRevLett.101.100501} {\bibfield  {journal} {\bibinfo  {journal}
  {Phys. Rev. Lett.}\ }\textbf {\bibinfo {volume} {101}},\ \bibinfo {pages}
  {100501} (\bibinfo {year} {2008}{\natexlab{b}})}\BibitemShut {NoStop}%
\bibitem [{\citenamefont {Zhou}\ \emph {et~al.}(2009)\citenamefont {Zhou},
  \citenamefont {Yang}, \citenamefont {Liu}, \citenamefont {Sun},\ and\
  \citenamefont {Nori}}]{PhysRevA.80.062109}%
  \BibitemOpen
  \bibfield  {author} {\bibinfo {author} {\bibfnamefont {L.}~\bibnamefont
  {Zhou}}, \bibinfo {author} {\bibfnamefont {S.}~\bibnamefont {Yang}}, \bibinfo
  {author} {\bibfnamefont {Y.-X.}\ \bibnamefont {Liu}}, \bibinfo {author}
  {\bibfnamefont {C.~P.}\ \bibnamefont {Sun}}, \ and\ \bibinfo {author}
  {\bibfnamefont {F.}~\bibnamefont {Nori}},\ }\bibfield  {title} {\emph
  {\bibinfo {title} {{Quantum Zeno switch for single-photon coherent
  transport}},\ }}\href {\doibase 10.1103/PhysRevA.80.062109} {\bibfield
  {journal} {\bibinfo  {journal} {Phys. Rev. A}\ }\textbf {\bibinfo {volume}
  {80}},\ \bibinfo {pages} {062109} (\bibinfo {year} {2009})}\BibitemShut
  {NoStop}%
\bibitem [{\citenamefont {Astafiev}\ \emph {et~al.}(2010)\citenamefont
  {Astafiev}, \citenamefont {Zagoskin}, \citenamefont {Abdumalikov},
  \citenamefont {Pashkin}, \citenamefont {Yamamoto}, \citenamefont {Inomata},
  \citenamefont {Nakamura},\ and\ \citenamefont
  {Tsai}}]{astafiev2010resonance}%
  \BibitemOpen
  \bibfield  {author} {\bibinfo {author} {\bibfnamefont {O.}~\bibnamefont
  {Astafiev}}, \bibinfo {author} {\bibfnamefont {A.~M.}\ \bibnamefont
  {Zagoskin}}, \bibinfo {author} {\bibfnamefont {A.~A.}\ \bibnamefont
  {Abdumalikov}}, \bibinfo {author} {\bibfnamefont {Y.~A.}\ \bibnamefont
  {Pashkin}}, \bibinfo {author} {\bibfnamefont {T.}~\bibnamefont {Yamamoto}},
  \bibinfo {author} {\bibfnamefont {K.}~\bibnamefont {Inomata}}, \bibinfo
  {author} {\bibfnamefont {Y.}~\bibnamefont {Nakamura}}, \ and\ \bibinfo
  {author} {\bibfnamefont {J.~S.}\ \bibnamefont {Tsai}},\ }\bibfield  {title}
  {\emph {\bibinfo {title} {{Resonance fluorescence of a single artificial
  atom}},\ }}\href {\doibase 10.1126/science.1181918} {\bibfield  {journal}
  {\bibinfo  {journal} {Science}\ }\textbf {\bibinfo {volume} {327}},\ \bibinfo
  {pages} {840} (\bibinfo {year} {2010})}\BibitemShut {NoStop}%
\bibitem [{\citenamefont {Peropadre}\ \emph {et~al.}(2013)\citenamefont
  {Peropadre}, \citenamefont {Lindkvist}, \citenamefont {Hoi}, \citenamefont
  {Wilson}, \citenamefont {Garcia-Ripoll}, \citenamefont {Delsing},\ and\
  \citenamefont {Johansson}}]{peropadre2013scattering}%
  \BibitemOpen
  \bibfield  {author} {\bibinfo {author} {\bibfnamefont {B.}~\bibnamefont
  {Peropadre}}, \bibinfo {author} {\bibfnamefont {J.}~\bibnamefont
  {Lindkvist}}, \bibinfo {author} {\bibfnamefont {I.-C.}\ \bibnamefont {Hoi}},
  \bibinfo {author} {\bibfnamefont {C. M.}~\bibnamefont {Wilson}}, \bibinfo
  {author} {\bibfnamefont {J.~J.}\ \bibnamefont {Garcia-Ripoll}}, \bibinfo
  {author} {\bibfnamefont {P.}~\bibnamefont {Delsing}}, \ and\ \bibinfo
  {author} {\bibfnamefont {G.}~\bibnamefont {Johansson}},\ }\bibfield  {title}
  {\emph {\bibinfo {title} {{Scattering of coherent states on a single
  artificial atom}},\ }}\href {\doibase 10.1088/1367-2630/15/3/035009}
  {\bibfield  {journal} {\bibinfo  {journal} {New J. Phys.}\ }\textbf {\bibinfo
  {volume} {15}},\ \bibinfo {pages} {035009} (\bibinfo {year}
  {2013})}\BibitemShut {NoStop}%
\bibitem [{\citenamefont {Van~Loo}\ \emph {et~al.}(2013)\citenamefont
  {Van~Loo}, \citenamefont {Fedorov}, \citenamefont {Lalumiere}, \citenamefont
  {Sanders}, \citenamefont {Blais},\ and\ \citenamefont
  {Wallraff}}]{van2013photon}%
  \BibitemOpen
  \bibfield  {author} {\bibinfo {author} {\bibfnamefont {A.~F.}\ \bibnamefont
  {Van~Loo}}, \bibinfo {author} {\bibfnamefont {A.}~\bibnamefont {Fedorov}},
  \bibinfo {author} {\bibfnamefont {K.}~\bibnamefont {Lalumiere}}, \bibinfo
  {author} {\bibfnamefont {B.~C.}\ \bibnamefont {Sanders}}, \bibinfo {author}
  {\bibfnamefont {A.}~\bibnamefont {Blais}}, \ and\ \bibinfo {author}
  {\bibfnamefont {A.}~\bibnamefont {Wallraff}},\ }\bibfield  {title} {\emph
  {\bibinfo {title} {{Photon-mediated interactions between distant artificial
  atoms}},\ }}\href {\doibase 10.1126/science.1244324} {\bibfield  {journal}
  {\bibinfo  {journal} {Science}\ }\textbf {\bibinfo {volume} {342}},\ \bibinfo
  {pages} {1494} (\bibinfo {year} {2013})}\BibitemShut {NoStop}%
\bibitem [{\citenamefont {Lodahl}\ \emph {et~al.}(2017)\citenamefont {Lodahl},
  \citenamefont {Mahmoodian}, \citenamefont {Stobbe}, \citenamefont
  {Rauschenbeutel}, \citenamefont {Schneeweiss}, \citenamefont {Volz},
  \citenamefont {Pichler},\ and\ \citenamefont {Zoller}}]{lodahl2017chiral}%
  \BibitemOpen
  \bibfield  {author} {\bibinfo {author} {\bibfnamefont {P.}~\bibnamefont
  {Lodahl}}, \bibinfo {author} {\bibfnamefont {S.}~\bibnamefont {Mahmoodian}},
  \bibinfo {author} {\bibfnamefont {S.}~\bibnamefont {Stobbe}}, \bibinfo
  {author} {\bibfnamefont {A.}~\bibnamefont {Rauschenbeutel}}, \bibinfo
  {author} {\bibfnamefont {P.}~\bibnamefont {Schneeweiss}}, \bibinfo {author}
  {\bibfnamefont {J.}~\bibnamefont {Volz}}, \bibinfo {author} {\bibfnamefont
  {H.}~\bibnamefont {Pichler}}, \ and\ \bibinfo {author} {\bibfnamefont
  {P.}~\bibnamefont {Zoller}},\ }\bibfield  {title} {\emph {\bibinfo {title}
  {Chiral quantum optics},\ }}\href {\doibase 10.1038/nature21037} {\bibfield
  {journal} {\bibinfo  {journal} {Nature}\ }\textbf {\bibinfo {volume} {541}},\
  \bibinfo {pages} {473} (\bibinfo {year} {2017})}\BibitemShut {NoStop}%
\bibitem [{\citenamefont {Mirhosseini}\ \emph {et~al.}(2019)\citenamefont
  {Mirhosseini}, \citenamefont {Kim}, \citenamefont {Zhang}, \citenamefont
  {Sipahigil}, \citenamefont {Dieterle}, \citenamefont {Keller}, \citenamefont
  {Asenjo-Garcia}, \citenamefont {Chang},\ and\ \citenamefont
  {Painter}}]{mirhosseini2019cavity}%
  \BibitemOpen
  \bibfield  {author} {\bibinfo {author} {\bibfnamefont {M.}~\bibnamefont
  {Mirhosseini}}, \bibinfo {author} {\bibfnamefont {E.}~\bibnamefont {Kim}},
  \bibinfo {author} {\bibfnamefont {X.}~\bibnamefont {Zhang}}, \bibinfo
  {author} {\bibfnamefont {A.}~\bibnamefont {Sipahigil}}, \bibinfo {author}
  {\bibfnamefont {P.~B.}\ \bibnamefont {Dieterle}}, \bibinfo {author}
  {\bibfnamefont {A.~J.}\ \bibnamefont {Keller}}, \bibinfo {author}
  {\bibfnamefont {A.}~\bibnamefont {Asenjo-Garcia}}, \bibinfo {author}
  {\bibfnamefont {D.~E.}\ \bibnamefont {Chang}}, \ and\ \bibinfo {author}
  {\bibfnamefont {O.}~\bibnamefont {Painter}},\ }\bibfield  {title} {\emph
  {\bibinfo {title} {{Cavity quantum electrodynamics with atom-like mirrors}},\
  }}\href {\doibase 10.1038/s41586-019-1196-1} {\bibfield  {journal} {\bibinfo
  {journal} {Nature}\ }\textbf {\bibinfo {volume} {569}},\ \bibinfo {pages}
  {692} (\bibinfo {year} {2019})}\BibitemShut {NoStop}%
\bibitem [{\citenamefont {Gong}\ \emph {et~al.}(2008)\citenamefont {Gong},
  \citenamefont {Ian}, \citenamefont {Zhou},\ and\ \citenamefont
  {Sun}}]{PhysRevA.78.053806}%
  \BibitemOpen
  \bibfield  {author} {\bibinfo {author} {\bibfnamefont {Z.~R.}\ \bibnamefont
  {Gong}}, \bibinfo {author} {\bibfnamefont {H.}~\bibnamefont {Ian}}, \bibinfo
  {author} {\bibfnamefont {L.}~\bibnamefont {Zhou}}, \ and\ \bibinfo {author}
  {\bibfnamefont {C.~P.}\ \bibnamefont {Sun}},\ }\bibfield  {title} {\emph
  {\bibinfo {title} {{Controlling quasibound states in a one-dimensional
  continuum through an electromagnetically-induced-transparency mechanism}},\
  }}\href {\doibase 10.1103/PhysRevA.78.053806} {\bibfield  {journal} {\bibinfo
   {journal} {Phys. Rev. A}\ }\textbf {\bibinfo {volume} {78}},\ \bibinfo
  {pages} {053806} (\bibinfo {year} {2008})}\BibitemShut {NoStop}%
\bibitem [{\citenamefont {Dong}\ \emph {et~al.}(2009)\citenamefont {Dong},
  \citenamefont {Gong}, \citenamefont {Ian}, \citenamefont {Zhou},\ and\
  \citenamefont {Sun}}]{PhysRevA.79.063847}%
  \BibitemOpen
  \bibfield  {author} {\bibinfo {author} {\bibfnamefont {H.}~\bibnamefont
  {Dong}}, \bibinfo {author} {\bibfnamefont {Z.~R.}\ \bibnamefont {Gong}},
  \bibinfo {author} {\bibfnamefont {H.}~\bibnamefont {Ian}}, \bibinfo {author}
  {\bibfnamefont {L.}~\bibnamefont {Zhou}}, \ and\ \bibinfo {author}
  {\bibfnamefont {C.~P.}\ \bibnamefont {Sun}},\ }\bibfield  {title} {\emph
  {\bibinfo {title} {{Intrinsic cavity QED and emergent quasinormal modes for a
  single photon}},\ }}\href {\doibase 10.1103/PhysRevA.79.063847} {\bibfield
  {journal} {\bibinfo  {journal} {Phys. Rev. A}\ }\textbf {\bibinfo {volume}
  {79}},\ \bibinfo {pages} {063847} (\bibinfo {year} {2009})}\BibitemShut
  {NoStop}%
\bibitem [{\citenamefont {Guimond}\ \emph {et~al.}(2016)\citenamefont
  {Guimond}, \citenamefont {Roulet}, \citenamefont {Le},\ and\ \citenamefont
  {Scarani}}]{PhysRevA.93.023808}%
  \BibitemOpen
  \bibfield  {author} {\bibinfo {author} {\bibfnamefont {P.-O.}\ \bibnamefont
  {Guimond}}, \bibinfo {author} {\bibfnamefont {A.}~\bibnamefont {Roulet}},
  \bibinfo {author} {\bibfnamefont {H.~N.}\ \bibnamefont {Le}}, \ and\ \bibinfo
  {author} {\bibfnamefont {V.}~\bibnamefont {Scarani}},\ }\bibfield  {title}
  {\emph {\bibinfo {title} {{Rabi oscillation in a quantum cavity: Markovian
  and non-Markovian dynamics}},\ }}\href {\doibase 10.1103/PhysRevA.93.023808}
  {\bibfield  {journal} {\bibinfo  {journal} {Phys. Rev. A}\ }\textbf {\bibinfo
  {volume} {93}},\ \bibinfo {pages} {023808} (\bibinfo {year}
  {2016})}\BibitemShut {NoStop}%
\bibitem [{\citenamefont {Calaj\'o}\ \emph {et~al.}(2019)\citenamefont
  {Calaj\'o}, \citenamefont {Fang}, \citenamefont {Baranger},\ and\
  \citenamefont {Ciccarello}}]{PhysRevLett.122.073601}%
  \BibitemOpen
  \bibfield  {author} {\bibinfo {author} {\bibfnamefont {G.}~\bibnamefont
  {Calaj\'o}}, \bibinfo {author} {\bibfnamefont {Y.-L.~L.}\ \bibnamefont
  {Fang}}, \bibinfo {author} {\bibfnamefont {H.~U.}\ \bibnamefont {Baranger}},
  \ and\ \bibinfo {author} {\bibfnamefont {F.}~\bibnamefont {Ciccarello}},\
  }\bibfield  {title} {\emph {\bibinfo {title} {{Exciting a Bound State in the
  Continuum through Multiphoton Scattering Plus Delayed Quantum Feedback}},\
  }}\href {\doibase 10.1103/PhysRevLett.122.073601} {\bibfield  {journal}
  {\bibinfo  {journal} {Phys. Rev. Lett.}\ }\textbf {\bibinfo {volume} {122}},\
  \bibinfo {pages} {073601} (\bibinfo {year} {2019})}\BibitemShut {NoStop}%
\bibitem [{\citenamefont {Song}\ \emph {et~al.}(2021)\citenamefont {Song},
  \citenamefont {Guo}, \citenamefont {Nie}, \citenamefont {Kwek},\ and\
  \citenamefont {Long}}]{song2021optical}%
  \BibitemOpen
  \bibfield  {author} {\bibinfo {author} {\bibfnamefont {G.-Z.}\ \bibnamefont
  {Song}}, \bibinfo {author} {\bibfnamefont {J.-L.}\ \bibnamefont {Guo}},
  \bibinfo {author} {\bibfnamefont {W.}~\bibnamefont {Nie}}, \bibinfo {author}
  {\bibfnamefont {L.-C.}\ \bibnamefont {Kwek}}, \ and\ \bibinfo {author}
  {\bibfnamefont {G.-L.}\ \bibnamefont {Long}},\ }\bibfield  {title} {\emph
  {\bibinfo {title} {{Optical properties of a waveguide-mediated chain of
  randomly positioned atoms}},\ }}\href {\doibase 10.1364/OE.409471} {\bibfield
   {journal} {\bibinfo  {journal} {Opt. Express}\ }\textbf {\bibinfo {volume}
  {29}},\ \bibinfo {pages} {1903} (\bibinfo {year} {2021})}\BibitemShut
  {NoStop}%
\bibitem [{\citenamefont {Arranz~Regidor}\ and\ \citenamefont
  {Hughes}(2021)}]{PhysRevA.104.L031701}%
  \BibitemOpen
  \bibfield  {author} {\bibinfo {author} {\bibfnamefont {S.}~\bibnamefont
  {Arranz~Regidor}}\ and\ \bibinfo {author} {\bibfnamefont {S.}~\bibnamefont
  {Hughes}},\ }\bibfield  {title} {\emph {\bibinfo {title} {{Cavitylike strong
  coupling in macroscopic waveguide QED using three coupled qubits in the deep
  non-Markovian regime}},\ }}\href {\doibase 10.1103/PhysRevA.104.L031701}
  {\bibfield  {journal} {\bibinfo  {journal} {Phys. Rev. A}\ }\textbf {\bibinfo
  {volume} {104}},\ \bibinfo {pages} {L031701} (\bibinfo {year}
  {2021})}\BibitemShut {NoStop}%
\bibitem [{\citenamefont {Zhou}\ \emph {et~al.}(2022)\citenamefont {Zhou},
  \citenamefont {Liao},\ and\ \citenamefont {Zubairy}}]{PhysRevA.105.033705}%
  \BibitemOpen
  \bibfield  {author} {\bibinfo {author} {\bibfnamefont {C.}~\bibnamefont
  {Zhou}}, \bibinfo {author} {\bibfnamefont {Z.}~\bibnamefont {Liao}}, \ and\
  \bibinfo {author} {\bibfnamefont {M.~S.}\ \bibnamefont {Zubairy}},\
  }\bibfield  {title} {\emph {\bibinfo {title} {{Decay of a single photon in a
  cavity with atomic mirrors}},\ }}\href {\doibase 10.1103/PhysRevA.105.033705}
  {\bibfield  {journal} {\bibinfo  {journal} {Phys. Rev. A}\ }\textbf {\bibinfo
  {volume} {105}},\ \bibinfo {pages} {033705} (\bibinfo {year}
  {2022})}\BibitemShut {NoStop}%
\bibitem [{\citenamefont {Dutra}(2005)}]{dutra2005cavity}%
  \BibitemOpen
  \bibfield  {author} {\bibinfo {author} {\bibfnamefont {S.~M.}\ \bibnamefont
  {Dutra}},\ }\href
  {https://www.wiley.com/en-us/Cavity+Quantum+Electrodynamics%3A+The+Strange+Theory+of+Light+in+a+Box+-p-9780471443384}
  {\emph {\bibinfo {title} {{Cavity Quantum Electrodynamics: The Strange Theory
  of Light in a Box}}}}\ (\bibinfo  {publisher} {Wiley, Hoboken, New Jersey},\
  \bibinfo {year} {2005})\BibitemShut {NoStop}%
\bibitem [{\citenamefont {Lamprecht}\ and\ \citenamefont
  {Ritsch}(1999)}]{PhysRevLett.82.3787}%
  \BibitemOpen
  \bibfield  {author} {\bibinfo {author} {\bibfnamefont {C.}~\bibnamefont
  {Lamprecht}}\ and\ \bibinfo {author} {\bibfnamefont {H.}~\bibnamefont
  {Ritsch}},\ }\bibfield  {title} {\emph {\bibinfo {title} {{Quantized
  Atom-Field Dynamics in Unstable Cavities}},\ }}\href {\doibase
  10.1103/PhysRevLett.82.3787} {\bibfield  {journal} {\bibinfo  {journal}
  {Phys. Rev. Lett.}\ }\textbf {\bibinfo {volume} {82}},\ \bibinfo {pages}
  {3787} (\bibinfo {year} {1999})}\BibitemShut {NoStop}%
\bibitem [{\citenamefont {Lalanne}\ \emph {et~al.}(2008)\citenamefont
  {Lalanne}, \citenamefont {Sauvan},\ and\ \citenamefont
  {Hugonin}}]{lalanne2008photon}%
  \BibitemOpen
  \bibfield  {author} {\bibinfo {author} {\bibfnamefont {P.}~\bibnamefont
  {Lalanne}}, \bibinfo {author} {\bibfnamefont {C.}~\bibnamefont {Sauvan}}, \
  and\ \bibinfo {author} {\bibfnamefont {J. P.}\ \bibnamefont {Hugonin}},\
  }\bibfield  {title} {\emph {\bibinfo {title} {{Photon confinement in photonic
  crystal nanocavities}},\ }}\href {\doibase 10.1002/lpor.200810018} {\bibfield
   {journal} {\bibinfo  {journal} {Laser Photonics Rev.}\ }\textbf {\bibinfo
  {volume} {2}},\ \bibinfo {pages} {514} (\bibinfo {year} {2008})}\BibitemShut
  {NoStop}%
\bibitem [{\citenamefont {H\"ummer}\ \emph {et~al.}(2013)\citenamefont
  {H\"ummer}, \citenamefont {Garc\'{\i}a-Vidal}, \citenamefont
  {Mart\'{\i}n-Moreno},\ and\ \citenamefont {Zueco}}]{PhysRevB.87.115419}%
  \BibitemOpen
  \bibfield  {author} {\bibinfo {author} {\bibfnamefont {T.}~\bibnamefont
  {H\"ummer}}, \bibinfo {author} {\bibfnamefont {F.~J.}\ \bibnamefont
  {Garc\'{\i}a-Vidal}}, \bibinfo {author} {\bibfnamefont {L.}~\bibnamefont
  {Mart\'{\i}n-Moreno}}, \ and\ \bibinfo {author} {\bibfnamefont
  {D.}~\bibnamefont {Zueco}},\ }\bibfield  {title} {\emph {\bibinfo {title}
  {{Weak and strong coupling regimes in plasmonic QED}},\ }}\href {\doibase
  10.1103/PhysRevB.87.115419} {\bibfield  {journal} {\bibinfo  {journal} {Phys.
  Rev. B}\ }\textbf {\bibinfo {volume} {87}},\ \bibinfo {pages} {115419}
  (\bibinfo {year} {2013})}\BibitemShut {NoStop}%
\bibitem [{\citenamefont {Kristensen}\ \emph {et~al.}(2015)\citenamefont
  {Kristensen}, \citenamefont {Ge},\ and\ \citenamefont
  {Hughes}}]{PhysRevA.92.053810}%
  \BibitemOpen
  \bibfield  {author} {\bibinfo {author} {\bibfnamefont {P.~T.}\ \bibnamefont
  {Kristensen}}, \bibinfo {author} {\bibfnamefont {R.-C.}\ \bibnamefont {Ge}},
  \ and\ \bibinfo {author} {\bibfnamefont {S.}~\bibnamefont {Hughes}},\
  }\bibfield  {title} {\emph {\bibinfo {title} {{Normalization of quasinormal
  modes in leaky optical cavities and plasmonic resonators}},\ }}\href
  {\doibase 10.1103/PhysRevA.92.053810} {\bibfield  {journal} {\bibinfo
  {journal} {Phys. Rev. A}\ }\textbf {\bibinfo {volume} {92}},\ \bibinfo
  {pages} {053810} (\bibinfo {year} {2015})}\BibitemShut {NoStop}%
\bibitem [{\citenamefont {Pellizzari}\ \emph {et~al.}(1995)\citenamefont
  {Pellizzari}, \citenamefont {Gardiner}, \citenamefont {Cirac},\ and\
  \citenamefont {Zoller}}]{PhysRevLett.75.3788}%
  \BibitemOpen
  \bibfield  {author} {\bibinfo {author} {\bibfnamefont {T.}~\bibnamefont
  {Pellizzari}}, \bibinfo {author} {\bibfnamefont {S.~A.}\ \bibnamefont
  {Gardiner}}, \bibinfo {author} {\bibfnamefont {J.~I.}\ \bibnamefont {Cirac}},
  \ and\ \bibinfo {author} {\bibfnamefont {P.}~\bibnamefont {Zoller}},\
  }\bibfield  {title} {\emph {\bibinfo {title} {{Decoherence, Continuous
  Observation, and Quantum Computing: A Cavity QED Model}},\ }}\href {\doibase
  10.1103/PhysRevLett.75.3788} {\bibfield  {journal} {\bibinfo  {journal}
  {Phys. Rev. Lett.}\ }\textbf {\bibinfo {volume} {75}},\ \bibinfo {pages}
  {3788} (\bibinfo {year} {1995})}\BibitemShut {NoStop}%
\bibitem [{\citenamefont {Gu}\ \emph {et~al.}(2017)\citenamefont {Gu},
  \citenamefont {Kockum}, \citenamefont {Miranowicz}, \citenamefont {Liu},\
  and\ \citenamefont {Nori}}]{gu2017microwave}%
  \BibitemOpen
  \bibfield  {author} {\bibinfo {author} {\bibfnamefont {X.}~\bibnamefont
  {Gu}}, \bibinfo {author} {\bibfnamefont {A.~F.}\ \bibnamefont {Kockum}},
  \bibinfo {author} {\bibfnamefont {A.}~\bibnamefont {Miranowicz}}, \bibinfo
  {author} {\bibfnamefont {Y.-X.}\ \bibnamefont {Liu}}, \ and\ \bibinfo
  {author} {\bibfnamefont {F.}~\bibnamefont {Nori}},\ }\bibfield  {title}
  {\emph {\bibinfo {title} {{Microwave photonics with superconducting quantum
  circuits}},\ }}\href {\doibase 10.1016/j.physrep.2017.10.002} {\bibfield
  {journal} {\bibinfo  {journal} {Phys. Rep.}\ }\textbf {\bibinfo {volume}
  {718-719}},\ \bibinfo {pages} {1} (\bibinfo {year} {2017})}\BibitemShut
  {NoStop}%
\bibitem [{\citenamefont {Kimble}(2008)}]{kimble2008quantum}%
  \BibitemOpen
  \bibfield  {author} {\bibinfo {author} {\bibfnamefont {H.~J.}\ \bibnamefont
  {Kimble}},\ }\bibfield  {title} {\emph {\bibinfo {title} {{The quantum
  internet}},\ }}\href {\doibase 10.1038/nature11023} {\bibfield  {journal}
  {\bibinfo  {journal} {Nature}\ }\textbf {\bibinfo {volume} {453}},\ \bibinfo
  {pages} {1023} (\bibinfo {year} {2008})}\BibitemShut {NoStop}%
\bibitem [{\citenamefont {Reiserer}\ and\ \citenamefont
  {Rempe}(2015)}]{RevModPhys.87.1379}%
  \BibitemOpen
  \bibfield  {author} {\bibinfo {author} {\bibfnamefont {A.}~\bibnamefont
  {Reiserer}}\ and\ \bibinfo {author} {\bibfnamefont {G.}~\bibnamefont
  {Rempe}},\ }\bibfield  {title} {\emph {\bibinfo {title} {{Cavity-based
  quantum networks with single atoms and optical photons}},\ }}\href {\doibase
  10.1103/RevModPhys.87.1379} {\bibfield  {journal} {\bibinfo  {journal} {Rev.
  Mod. Phys.}\ }\textbf {\bibinfo {volume} {87}},\ \bibinfo {pages} {1379}
  (\bibinfo {year} {2015})}\BibitemShut {NoStop}%
\bibitem [{\citenamefont {Tian}\ \emph {et~al.}(2021)\citenamefont {Tian},
  \citenamefont {Zhang},\ and\ \citenamefont
  {Chen}}]{PhysRevApplied.15.054043}%
  \BibitemOpen
  \bibfield  {author} {\bibinfo {author} {\bibfnamefont {Z.}~\bibnamefont
  {Tian}}, \bibinfo {author} {\bibfnamefont {P.}~\bibnamefont {Zhang}}, \ and\
  \bibinfo {author} {\bibfnamefont {X.-W.}\ \bibnamefont {Chen}},\ }\bibfield
  {title} {\emph {\bibinfo {title} {{Static Hybrid Quantum Nodes: Toward
  Perfect State Transfer on a Photonic Chip}},\ }}\href {\doibase
  10.1103/PhysRevApplied.15.054043} {\bibfield  {journal} {\bibinfo  {journal}
  {Phys. Rev. Appl.}\ }\textbf {\bibinfo {volume} {15}},\ \bibinfo {pages}
  {054043} (\bibinfo {year} {2021})}\BibitemShut {NoStop}%
\bibitem [{\citenamefont {Aspelmeyer}\ \emph {et~al.}(2014)\citenamefont
  {Aspelmeyer}, \citenamefont {Kippenberg},\ and\ \citenamefont
  {Marquardt}}]{RevModPhys.86.1391}%
  \BibitemOpen
  \bibfield  {author} {\bibinfo {author} {\bibfnamefont {M.}~\bibnamefont
  {Aspelmeyer}}, \bibinfo {author} {\bibfnamefont {T.~J.}\ \bibnamefont
  {Kippenberg}}, \ and\ \bibinfo {author} {\bibfnamefont {F.}~\bibnamefont
  {Marquardt}},\ }\bibfield  {title} {\emph {\bibinfo {title} {{Cavity
  optomechanics}},\ }}\href {\doibase 10.1103/RevModPhys.86.1391} {\bibfield
  {journal} {\bibinfo  {journal} {Rev. Mod. Phys.}\ }\textbf {\bibinfo {volume}
  {86}},\ \bibinfo {pages} {1391} (\bibinfo {year} {2014})}\BibitemShut
  {NoStop}%
\bibitem [{\citenamefont {Li}\ and\ \citenamefont
  {Wei}(2015)}]{PhysRevA.92.063836}%
  \BibitemOpen
  \bibfield  {author} {\bibinfo {author} {\bibfnamefont {X.}~\bibnamefont
  {Li}}\ and\ \bibinfo {author} {\bibfnamefont {L.~F.}\ \bibnamefont {Wei}},\
  }\bibfield  {title} {\emph {\bibinfo {title} {Designable single-photon
  quantum routings with atomic mirrors},\ }}\href {\doibase
  10.1103/PhysRevA.92.063836} {\bibfield  {journal} {\bibinfo  {journal} {Phys.
  Rev. A}\ }\textbf {\bibinfo {volume} {92}},\ \bibinfo {pages} {063836}
  (\bibinfo {year} {2015})}\BibitemShut {NoStop}%
\bibitem [{\citenamefont {Qian}\ \emph {et~al.}(2021)\citenamefont {Qian},
  \citenamefont {Shan}, \citenamefont {Zhang}, \citenamefont {Liu},
  \citenamefont {Ma}, \citenamefont {Gong},\ and\ \citenamefont
  {Gu}}]{qian2021spontaneous}%
  \BibitemOpen
  \bibfield  {author} {\bibinfo {author} {\bibfnamefont {Z.}~\bibnamefont
  {Qian}}, \bibinfo {author} {\bibfnamefont {L.}~\bibnamefont {Shan}}, \bibinfo
  {author} {\bibfnamefont {X.}~\bibnamefont {Zhang}}, \bibinfo {author}
  {\bibfnamefont {Q.}~\bibnamefont {Liu}}, \bibinfo {author} {\bibfnamefont
  {Y.}~\bibnamefont {Ma}}, \bibinfo {author} {\bibfnamefont {Q.}~\bibnamefont
  {Gong}}, \ and\ \bibinfo {author} {\bibfnamefont {Y.}~\bibnamefont {Gu}},\
  }\bibfield  {title} {\emph {\bibinfo {title} {{Spontaneous emission in micro-
  or nanophotonic structures}},\ }}\href {\doibase 10.1186/s43074-021-00043-z}
  {\bibfield  {journal} {\bibinfo  {journal} {PhotoniX}\ }\textbf {\bibinfo
  {volume} {2}},\ \bibinfo {pages} {21} (\bibinfo {year} {2021})}\BibitemShut
  {NoStop}%
\bibitem [{\citenamefont {Friedler}\ \emph {et~al.}(2009)\citenamefont
  {Friedler}, \citenamefont {Sauvan}, \citenamefont {Hugonin}, \citenamefont
  {Lalanne}, \citenamefont {Claudon},\ and\ \citenamefont
  {G{\'e}rard}}]{Friedler2009solid}%
  \BibitemOpen
  \bibfield  {author} {\bibinfo {author} {\bibfnamefont {I.}~\bibnamefont
  {Friedler}}, \bibinfo {author} {\bibfnamefont {C.}~\bibnamefont {Sauvan}},
  \bibinfo {author} {\bibfnamefont {J.-P.}\ \bibnamefont {Hugonin}}, \bibinfo
  {author} {\bibfnamefont {P.}~\bibnamefont {Lalanne}}, \bibinfo {author}
  {\bibfnamefont {J.}~\bibnamefont {Claudon}}, \ and\ \bibinfo {author}
  {\bibfnamefont {J. M.}\ \bibnamefont {G{\'e}rard}},\ }\bibfield  {title}
  {\emph {\bibinfo {title} {{Solid-state single photon sources: the nanowire
  antenna}},\ }}\href {\doibase 10.1364/OE.17.002095} {\bibfield  {journal}
  {\bibinfo  {journal} {Opt. Express}\ }\textbf {\bibinfo {volume} {17}},\
  \bibinfo {pages} {2095} (\bibinfo {year} {2009})}\BibitemShut {NoStop}%
\bibitem [{\citenamefont {Denning}\ \emph {et~al.}(2018)\citenamefont
  {Denning}, \citenamefont {Iles-Smith}, \citenamefont {Osterkryger},
  \citenamefont {Gregersen},\ and\ \citenamefont {Mork}}]{PhysRevB.98.121306}%
  \BibitemOpen
  \bibfield  {author} {\bibinfo {author} {\bibfnamefont {E.~V.}\ \bibnamefont
  {Denning}}, \bibinfo {author} {\bibfnamefont {J.}~\bibnamefont {Iles-Smith}},
  \bibinfo {author} {\bibfnamefont {A.~D.}\ \bibnamefont {Osterkryger}},
  \bibinfo {author} {\bibfnamefont {N.}~\bibnamefont {Gregersen}}, \ and\
  \bibinfo {author} {\bibfnamefont {J.}~\bibnamefont {Mork}},\ }\bibfield
  {title} {\emph {\bibinfo {title} {{Cavity-waveguide interplay in optical
  resonators and its role in optimal single-photon sources}},\ }}\href
  {\doibase 10.1103/PhysRevB.98.121306} {\bibfield  {journal} {\bibinfo
  {journal} {Phys. Rev. B}\ }\textbf {\bibinfo {volume} {98}},\ \bibinfo
  {pages} {121306(R)} (\bibinfo {year} {2018})}\BibitemShut {NoStop}%
\bibitem [{\citenamefont {Mork}\ \emph {et~al.}(2014)\citenamefont {Mork},
  \citenamefont {Chen},\ and\ \citenamefont {Heuck}}]{PhysRevLett.113.163901}%
  \BibitemOpen
  \bibfield  {author} {\bibinfo {author} {\bibfnamefont {J.}~\bibnamefont
  {Mork}}, \bibinfo {author} {\bibfnamefont {Y.}~\bibnamefont {Chen}}, \ and\
  \bibinfo {author} {\bibfnamefont {M.}~\bibnamefont {Heuck}},\ }\bibfield
  {title} {\emph {\bibinfo {title} {{Photonic Crystal Fano Laser: Terahertz
  Modulation and Ultrashort Pulse Generation}},\ }}\href {\doibase
  10.1103/PhysRevLett.113.163901} {\bibfield  {journal} {\bibinfo  {journal}
  {Phys. Rev. Lett.}\ }\textbf {\bibinfo {volume} {113}},\ \bibinfo {pages}
  {163901} (\bibinfo {year} {2014})}\BibitemShut {NoStop}%
\bibitem [{\citenamefont {Yu}\ \emph {et~al.}(2017)\citenamefont {Yu},
  \citenamefont {Xue}, \citenamefont {Semenova}, \citenamefont {Yvind},\ and\
  \citenamefont {Mork}}]{yu2017demonstration}%
  \BibitemOpen
  \bibfield  {author} {\bibinfo {author} {\bibfnamefont {Y.}~\bibnamefont
  {Yu}}, \bibinfo {author} {\bibfnamefont {W.}~\bibnamefont {Xue}}, \bibinfo
  {author} {\bibfnamefont {E.}~\bibnamefont {Semenova}}, \bibinfo {author}
  {\bibfnamefont {K.}~\bibnamefont {Yvind}}, \ and\ \bibinfo {author}
  {\bibfnamefont {J.}~\bibnamefont {Mork}},\ }\bibfield  {title} {\emph
  {\bibinfo {title} {{Demonstration of a self-pulsing photonic crystal Fano
  laser}},\ }}\href {\doibase 10.1038/nphoton.2016.248} {\bibfield  {journal}
  {\bibinfo  {journal} {Nature Photon.}\ }\textbf {\bibinfo {volume} {11}},\
  \bibinfo {pages} {81} (\bibinfo {year} {2017})}\BibitemShut {NoStop}%
\bibitem [{\citenamefont {Chang}\ \emph {et~al.}(2018)\citenamefont {Chang},
  \citenamefont {Douglas}, \citenamefont {Gonz\'alez-Tudela}, \citenamefont
  {Hung},\ and\ \citenamefont {Kimble}}]{RevModPhys.90.031002}%
  \BibitemOpen
  \bibfield  {author} {\bibinfo {author} {\bibfnamefont {D.~E.}\ \bibnamefont
  {Chang}}, \bibinfo {author} {\bibfnamefont {J.~S.}\ \bibnamefont {Douglas}},
  \bibinfo {author} {\bibfnamefont {A.}~\bibnamefont {Gonz\'alez-Tudela}},
  \bibinfo {author} {\bibfnamefont {C.-L.}\ \bibnamefont {Hung}}, \ and\
  \bibinfo {author} {\bibfnamefont {H.~J.}\ \bibnamefont {Kimble}},\ }\bibfield
   {title} {\emph {\bibinfo {title} {{Colloquium: Quantum matter built from
  nanoscopic lattices of atoms and photons}},\ }}\href {\doibase
  10.1103/RevModPhys.90.031002} {\bibfield  {journal} {\bibinfo  {journal}
  {Rev. Mod. Phys.}\ }\textbf {\bibinfo {volume} {90}},\ \bibinfo {pages}
  {031002} (\bibinfo {year} {2018})}\BibitemShut {NoStop}%
\bibitem [{\citenamefont {Sheremet}\ \emph {et~al.}(2023)\citenamefont
  {Sheremet}, \citenamefont {Petrov}, \citenamefont {Iorsh}, \citenamefont
  {Poshakinskiy},\ and\ \citenamefont {Poddubny}}]{RevModPhys.95.015002}%
  \BibitemOpen
  \bibfield  {author} {\bibinfo {author} {\bibfnamefont {A.~S.}\ \bibnamefont
  {Sheremet}}, \bibinfo {author} {\bibfnamefont {M.~I.}\ \bibnamefont
  {Petrov}}, \bibinfo {author} {\bibfnamefont {I.~V.}\ \bibnamefont {Iorsh}},
  \bibinfo {author} {\bibfnamefont {A.~V.}\ \bibnamefont {Poshakinskiy}}, \
  and\ \bibinfo {author} {\bibfnamefont {A.~N.}\ \bibnamefont {Poddubny}},\
  }\bibfield  {title} {\emph {\bibinfo {title} {{Waveguide quantum
  electrodynamics: Collective radiance and photon-photon correlations}},\
  }}\href {\doibase 10.1103/RevModPhys.95.015002} {\bibfield  {journal}
  {\bibinfo  {journal} {Rev. Mod. Phys.}\ }\textbf {\bibinfo {volume} {95}},\
  \bibinfo {pages} {015002} (\bibinfo {year} {2023})}\BibitemShut {NoStop}%
\bibitem [{\citenamefont {Kannan}\ \emph {et~al.}(2023)\citenamefont {Kannan},
  \citenamefont {Almanakly}, \citenamefont {Sung}, \citenamefont {Di~Paolo},
  \citenamefont {Rower}, \citenamefont {Braum{\"u}ller}, \citenamefont
  {Melville}, \citenamefont {Niedzielski}, \citenamefont {Karamlou},
  \citenamefont {Serniak}, \citenamefont {Veps{\"a}l{\"a}inen}, \citenamefont
  {Schwartz}, \citenamefont {Yoder}, \citenamefont {Winik}, \citenamefont
  {Wang}, \citenamefont {Orlando}, \citenamefont {Gustavsson}, \citenamefont
  {Grover},\ and\ \citenamefont {Oliver}}]{kannan2023}%
  \BibitemOpen
  \bibfield  {author} {\bibinfo {author} {\bibfnamefont {B.}~\bibnamefont
  {Kannan}}, \bibinfo {author} {\bibfnamefont {A.}~\bibnamefont {Almanakly}},
  \bibinfo {author} {\bibfnamefont {Y.}~\bibnamefont {Sung}}, \bibinfo {author}
  {\bibfnamefont {A.}~\bibnamefont {Di~Paolo}}, \bibinfo {author}
  {\bibfnamefont {D.~A.}\ \bibnamefont {Rower}}, \bibinfo {author}
  {\bibfnamefont {J.}~\bibnamefont {Braum{\"u}ller}}, \bibinfo {author}
  {\bibfnamefont {A.}~\bibnamefont {Melville}}, \bibinfo {author}
  {\bibfnamefont {B.~M.}\ \bibnamefont {Niedzielski}}, \bibinfo {author}
  {\bibfnamefont {A.}~\bibnamefont {Karamlou}}, \bibinfo {author}
  {\bibfnamefont {K.}~\bibnamefont {Serniak}}, \bibinfo {author} {\bibfnamefont
  {A.}~\bibnamefont {Veps{\"a}l{\"a}inen}}, \bibinfo {author} {\bibfnamefont
  {M.~E.}\ \bibnamefont {Schwartz}}, \bibinfo {author} {\bibfnamefont {J.~L.}\
  \bibnamefont {Yoder}}, \bibinfo {author} {\bibfnamefont {R.}~\bibnamefont
  {Winik}}, \bibinfo {author} {\bibfnamefont {J.~I.-J.}\ \bibnamefont {Wang}},
  \bibinfo {author} {\bibfnamefont {T.~P.}\ \bibnamefont {Orlando}}, \bibinfo
  {author} {\bibfnamefont {S.}~\bibnamefont {Gustavsson}}, \bibinfo {author}
  {\bibfnamefont {J.~A.}\ \bibnamefont {Grover}}, \ and\ \bibinfo {author}
  {\bibfnamefont {W.~D.}\ \bibnamefont {Oliver}},\ }\bibfield  {title} {\emph
  {\bibinfo {title} {{On-demand directional microwave photon emission using
  waveguide quantum electrodynamics}},\ }}\href {\doibase
  10.1038/s41567-022-01869-5} {\bibfield  {journal} {\bibinfo  {journal} {Nat.
  Phys.}\ }\textbf {\bibinfo {volume} {19}},\ \bibinfo {pages} {394} (\bibinfo
  {year} {2023})}\BibitemShut {NoStop}%
\bibitem [{\citenamefont {Peng}\ \emph {et~al.}(2016)\citenamefont {Peng},
  \citenamefont {Cao}, \citenamefont {Shen}, \citenamefont {Qu}, \citenamefont
  {Wen}, \citenamefont {Jiang},\ and\ \citenamefont {Xiao}}]{peng2016anti}%
  \BibitemOpen
  \bibfield  {author} {\bibinfo {author} {\bibfnamefont {P.}~\bibnamefont
  {Peng}}, \bibinfo {author} {\bibfnamefont {W.}~\bibnamefont {Cao}}, \bibinfo
  {author} {\bibfnamefont {C.}~\bibnamefont {Shen}}, \bibinfo {author}
  {\bibfnamefont {W.}~\bibnamefont {Qu}}, \bibinfo {author} {\bibfnamefont
  {J.}~\bibnamefont {Wen}}, \bibinfo {author} {\bibfnamefont {L.}~\bibnamefont
  {Jiang}}, \ and\ \bibinfo {author} {\bibfnamefont {Y.}~\bibnamefont {Xiao}},\
  }\bibfield  {title} {\emph {\bibinfo {title} {{Anti-parity-time symmetry with
  flying atoms}},\ }}\href {\doibase 10.1038/nphys3842} {\bibfield  {journal}
  {\bibinfo  {journal} {Nat. Phys.}\ }\textbf {\bibinfo {volume} {12}},\
  \bibinfo {pages} {1139} (\bibinfo {year} {2016})}\BibitemShut {NoStop}%
\bibitem [{\citenamefont {Choi}\ \emph {et~al.}(2018)\citenamefont {Choi},
  \citenamefont {Hahn}, \citenamefont {Yoon},\ and\ \citenamefont
  {Song}}]{choi2018observation}%
  \BibitemOpen
  \bibfield  {author} {\bibinfo {author} {\bibfnamefont {Y.}~\bibnamefont
  {Choi}}, \bibinfo {author} {\bibfnamefont {C.}~\bibnamefont {Hahn}}, \bibinfo
  {author} {\bibfnamefont {J.~W.}\ \bibnamefont {Yoon}}, \ and\ \bibinfo
  {author} {\bibfnamefont {S.~H.}\ \bibnamefont {Song}},\ }\bibfield  {title}
  {\emph {\bibinfo {title} {{Observation of an anti-PT-symmetric exceptional
  point and energy-difference conserving dynamics in electrical circuit
  resonators}},\ }}\href {\doibase 10.1038/s41467-018-04690-y} {\bibfield
  {journal} {\bibinfo  {journal} {Nat. Commun.}\ }\textbf {\bibinfo {volume}
  {9}},\ \bibinfo {pages} {2182} (\bibinfo {year} {2018})}\BibitemShut
  {NoStop}%
\bibitem [{\citenamefont {Li}\ \emph {et~al.}(2019{\natexlab{a}})\citenamefont
  {Li}, \citenamefont {Peng}, \citenamefont {Han}, \citenamefont {Miri},
  \citenamefont {Li}, \citenamefont {Xiao}, \citenamefont {Zhu}, \citenamefont
  {Zhao}, \citenamefont {Al{\`u}}, \citenamefont {Fan},\ and\ \citenamefont
  {Qiu}}]{li2019anti}%
  \BibitemOpen
  \bibfield  {author} {\bibinfo {author} {\bibfnamefont {Y.}~\bibnamefont
  {Li}}, \bibinfo {author} {\bibfnamefont {Y.-G.}\ \bibnamefont {Peng}},
  \bibinfo {author} {\bibfnamefont {L.}~\bibnamefont {Han}}, \bibinfo {author}
  {\bibfnamefont {M.-A.}\ \bibnamefont {Miri}}, \bibinfo {author}
  {\bibfnamefont {W.}~\bibnamefont {Li}}, \bibinfo {author} {\bibfnamefont
  {M.}~\bibnamefont {Xiao}}, \bibinfo {author} {\bibfnamefont {X.-F.}\
  \bibnamefont {Zhu}}, \bibinfo {author} {\bibfnamefont {J.}~\bibnamefont
  {Zhao}}, \bibinfo {author} {\bibfnamefont {A.}~\bibnamefont {Al{\`u}}},
  \bibinfo {author} {\bibfnamefont {S.}~\bibnamefont {Fan}}, \ and\ \bibinfo
  {author} {\bibfnamefont {C.-W.}\ \bibnamefont {Qiu}},\ }\bibfield  {title}
  {\emph {\bibinfo {title} {{Anti-parity-time symmetry in diffusive systems}},\
  }}\href {\doibase 10.1126/science.aaw6259} {\bibfield  {journal} {\bibinfo
  {journal} {Science}\ }\textbf {\bibinfo {volume} {364}},\ \bibinfo {pages}
  {170} (\bibinfo {year} {2019}{\natexlab{a}})}\BibitemShut {NoStop}%
\bibitem [{\citenamefont {Jiang}\ \emph {et~al.}(2019)\citenamefont {Jiang},
  \citenamefont {Mei}, \citenamefont {Zuo}, \citenamefont {Zhai}, \citenamefont
  {Li}, \citenamefont {Wen},\ and\ \citenamefont
  {Du}}]{PhysRevLett.123.193604}%
  \BibitemOpen
  \bibfield  {author} {\bibinfo {author} {\bibfnamefont {Y.}~\bibnamefont
  {Jiang}}, \bibinfo {author} {\bibfnamefont {Y.}~\bibnamefont {Mei}}, \bibinfo
  {author} {\bibfnamefont {Y.}~\bibnamefont {Zuo}}, \bibinfo {author}
  {\bibfnamefont {Y.}~\bibnamefont {Zhai}}, \bibinfo {author} {\bibfnamefont
  {J.}~\bibnamefont {Li}}, \bibinfo {author} {\bibfnamefont {J.}~\bibnamefont
  {Wen}}, \ and\ \bibinfo {author} {\bibfnamefont {S.}~\bibnamefont {Du}},\
  }\bibfield  {title} {\emph {\bibinfo {title} {{Anti-Parity-Time Symmetric
  Optical Four-Wave Mixing in Cold Atoms}},\ }}\href {\doibase
  10.1103/PhysRevLett.123.193604} {\bibfield  {journal} {\bibinfo  {journal}
  {Phys. Rev. Lett.}\ }\textbf {\bibinfo {volume} {123}},\ \bibinfo {pages}
  {193604} (\bibinfo {year} {2019})}\BibitemShut {NoStop}%
\bibitem [{\citenamefont {Zhao}\ \emph {et~al.}(2020)\citenamefont {Zhao},
  \citenamefont {Liu}, \citenamefont {Wu}, \citenamefont {Duan}, \citenamefont
  {Liu},\ and\ \citenamefont {Du}}]{PhysRevApplied.13.014053}%
  \BibitemOpen
  \bibfield  {author} {\bibinfo {author} {\bibfnamefont {J.}~\bibnamefont
  {Zhao}}, \bibinfo {author} {\bibfnamefont {Y.}~\bibnamefont {Liu}}, \bibinfo
  {author} {\bibfnamefont {L.}~\bibnamefont {Wu}}, \bibinfo {author}
  {\bibfnamefont {C.-K.}\ \bibnamefont {Duan}}, \bibinfo {author}
  {\bibfnamefont {Y.-X.}\ \bibnamefont {Liu}}, \ and\ \bibinfo {author}
  {\bibfnamefont {J.}~\bibnamefont {Du}},\ }\bibfield  {title} {\emph {\bibinfo
  {title} {{Observation of Anti-$\mathcal{P}\mathcal{T}$-Symmetry Phase
  Transition in the Magnon-Cavity-Magnon Coupled System}},\ }}\href {\doibase
  10.1103/PhysRevApplied.13.014053} {\bibfield  {journal} {\bibinfo  {journal}
  {Phys. Rev. Appl.}\ }\textbf {\bibinfo {volume} {13}},\ \bibinfo {pages}
  {014053} (\bibinfo {year} {2020})}\BibitemShut {NoStop}%
\bibitem [{\citenamefont {Zhang}\ \emph {et~al.}(2020)\citenamefont {Zhang},
  \citenamefont {Huang}, \citenamefont {Zhang}, \citenamefont {Li},
  \citenamefont {Qiu}, \citenamefont {Nori},\ and\ \citenamefont
  {Jing}}]{zhang2020breaking}%
  \BibitemOpen
  \bibfield  {author} {\bibinfo {author} {\bibfnamefont {H.}~\bibnamefont
  {Zhang}}, \bibinfo {author} {\bibfnamefont {R.}~\bibnamefont {Huang}},
  \bibinfo {author} {\bibfnamefont {S.-D.}\ \bibnamefont {Zhang}}, \bibinfo
  {author} {\bibfnamefont {Y.}~\bibnamefont {Li}}, \bibinfo {author}
  {\bibfnamefont {C.-W.}\ \bibnamefont {Qiu}}, \bibinfo {author} {\bibfnamefont
  {F.}~\bibnamefont {Nori}}, \ and\ \bibinfo {author} {\bibfnamefont
  {H.}~\bibnamefont {Jing}},\ }\bibfield  {title} {\emph {\bibinfo {title}
  {{Breaking anti-PT symmetry by spinning a resonator}},\ }}\href {\doibase
  10.1021/acs.nanolett.0c03119} {\bibfield  {journal} {\bibinfo  {journal}
  {Nano Lett.}\ }\textbf {\bibinfo {volume} {20}},\ \bibinfo {pages} {7594}
  (\bibinfo {year} {2020})}\BibitemShut {NoStop}%
\bibitem [{\citenamefont {Yang}\ \emph {et~al.}(2020)\citenamefont {Yang},
  \citenamefont {Wang}, \citenamefont {Rao}, \citenamefont {Gui}, \citenamefont
  {Yao}, \citenamefont {Lu},\ and\ \citenamefont
  {Hu}}]{PhysRevLett.125.147202}%
  \BibitemOpen
  \bibfield  {author} {\bibinfo {author} {\bibfnamefont {Y.}~\bibnamefont
  {Yang}}, \bibinfo {author} {\bibfnamefont {Y.-P.}\ \bibnamefont {Wang}},
  \bibinfo {author} {\bibfnamefont {J.~W.}\ \bibnamefont {Rao}}, \bibinfo
  {author} {\bibfnamefont {Y.~S.}\ \bibnamefont {Gui}}, \bibinfo {author}
  {\bibfnamefont {B.~M.}\ \bibnamefont {Yao}}, \bibinfo {author} {\bibfnamefont
  {W.}~\bibnamefont {Lu}}, \ and\ \bibinfo {author} {\bibfnamefont {C.-M.}\
  \bibnamefont {Hu}},\ }\bibfield  {title} {\emph {\bibinfo {title}
  {{Unconventional Singularity in Anti-Parity-Time Symmetric Cavity
  Magnonics}},\ }}\href {\doibase 10.1103/PhysRevLett.125.147202} {\bibfield
  {journal} {\bibinfo  {journal} {Phys. Rev. Lett.}\ }\textbf {\bibinfo
  {volume} {125}},\ \bibinfo {pages} {147202} (\bibinfo {year}
  {2020})}\BibitemShut {NoStop}%
\bibitem [{\citenamefont {Nair}\ \emph {et~al.}(2021)\citenamefont {Nair},
  \citenamefont {Mukhopadhyay},\ and\ \citenamefont
  {Agarwal}}]{PhysRevLett.126.180401}%
  \BibitemOpen
  \bibfield  {author} {\bibinfo {author} {\bibfnamefont {J.~M.~P.}\
  \bibnamefont {Nair}}, \bibinfo {author} {\bibfnamefont {D.}~\bibnamefont
  {Mukhopadhyay}}, \ and\ \bibinfo {author} {\bibfnamefont {G.~S.}\
  \bibnamefont {Agarwal}},\ }\bibfield  {title} {\emph {\bibinfo {title}
  {{Enhanced Sensing of Weak Anharmonicities through Coherences in
  Dissipatively Coupled Anti-PT Symmetric Systems}},\ }}\href {\doibase
  10.1103/PhysRevLett.126.180401} {\bibfield  {journal} {\bibinfo  {journal}
  {Phys. Rev. Lett.}\ }\textbf {\bibinfo {volume} {126}},\ \bibinfo {pages}
  {180401} (\bibinfo {year} {2021})}\BibitemShut {NoStop}%
\bibitem [{\citenamefont {Yang}\ \emph {et~al.}(2022)\citenamefont {Yang},
  \citenamefont {Xie}, \citenamefont {Li}, \citenamefont {Zhang}, \citenamefont
  {Peng}, \citenamefont {Wang}, \citenamefont {Li}, \citenamefont {Li},
  \citenamefont {Chen},\ and\ \citenamefont {Gao}}]{yang2022radiative}%
  \BibitemOpen
  \bibfield  {author} {\bibinfo {author} {\bibfnamefont {Y.}~\bibnamefont
  {Yang}}, \bibinfo {author} {\bibfnamefont {X.}~\bibnamefont {Xie}}, \bibinfo
  {author} {\bibfnamefont {Y.}~\bibnamefont {Li}}, \bibinfo {author}
  {\bibfnamefont {Z.}~\bibnamefont {Zhang}}, \bibinfo {author} {\bibfnamefont
  {Y.}~\bibnamefont {Peng}}, \bibinfo {author} {\bibfnamefont {C.}~\bibnamefont
  {Wang}}, \bibinfo {author} {\bibfnamefont {E.}~\bibnamefont {Li}}, \bibinfo
  {author} {\bibfnamefont {Y.}~\bibnamefont {Li}}, \bibinfo {author}
  {\bibfnamefont {H.}~\bibnamefont {Chen}}, \ and\ \bibinfo {author}
  {\bibfnamefont {F.}~\bibnamefont {Gao}},\ }\bibfield  {title} {\emph
  {\bibinfo {title} {{Radiative anti-parity-time plasmonics}},\ }}\href
  {\doibase 10.1038/s41467-022-35447-3} {\bibfield  {journal} {\bibinfo
  {journal} {Nat. Commun.}\ }\textbf {\bibinfo {volume} {13}},\ \bibinfo
  {pages} {7678} (\bibinfo {year} {2022})}\BibitemShut {NoStop}%
\bibitem [{\citenamefont {Tiranov}\ \emph {et~al.}(2023)\citenamefont
  {Tiranov}, \citenamefont {Angelopoulou}, \citenamefont {Jacobus~van Diepen},
  \citenamefont {Schrinski}, \citenamefont {Sandberg}, \citenamefont {Wang},
  \citenamefont {Midolo}, \citenamefont {Scholz}, \citenamefont {Wieck},
  \citenamefont {Ludwig}, \citenamefont {S{\o}rensen},\ and\ \citenamefont
  {Lodahl}}]{tiranov2023}%
  \BibitemOpen
  \bibfield  {author} {\bibinfo {author} {\bibfnamefont {A.}~\bibnamefont
  {Tiranov}}, \bibinfo {author} {\bibfnamefont {V.}~\bibnamefont
  {Angelopoulou}}, \bibinfo {author} {\bibfnamefont {C. J.}~\bibnamefont
  {van Diepen}}, \bibinfo {author} {\bibfnamefont {B.}~\bibnamefont
  {Schrinski}}, \bibinfo {author} {\bibfnamefont {O.~A.~D.}\ \bibnamefont
  {Sandberg}}, \bibinfo {author} {\bibfnamefont {Y.}~\bibnamefont {Wang}},
  \bibinfo {author} {\bibfnamefont {L.}~\bibnamefont {Midolo}}, \bibinfo
  {author} {\bibfnamefont {S.}~\bibnamefont {Scholz}}, \bibinfo {author}
  {\bibfnamefont {A.~D.}\ \bibnamefont {Wieck}}, \bibinfo {author}
  {\bibfnamefont {A.}~\bibnamefont {Ludwig}}, \bibinfo {author} {\bibfnamefont
  {A.~S.}\ \bibnamefont {S{\o}rensen}}, \ and\ \bibinfo {author} {\bibfnamefont
  {P.}~\bibnamefont {Lodahl}},\ }\bibfield  {title} {\emph {\bibinfo {title}
  {Collective super- and subradiant dynamics between distant optical quantum
  emitters},\ }}\href {\doibase 10.1126/science.ade9324} {\bibfield  {journal}
  {\bibinfo  {journal} {Science}\ }\textbf {\bibinfo {volume} {379}},\ \bibinfo
  {pages} {389} (\bibinfo {year} {2023})}\BibitemShut {NoStop}%
\bibitem [{\citenamefont {Corzo}\ \emph {et~al.}(2016)\citenamefont {Corzo},
  \citenamefont {Gouraud}, \citenamefont {Chandra}, \citenamefont {Goban},
  \citenamefont {Sheremet}, \citenamefont {Kupriyanov},\ and\ \citenamefont
  {Laurat}}]{PhysRevLett.117.133603}%
  \BibitemOpen
  \bibfield  {author} {\bibinfo {author} {\bibfnamefont {N.~V.}\ \bibnamefont
  {Corzo}}, \bibinfo {author} {\bibfnamefont {B.}~\bibnamefont {Gouraud}},
  \bibinfo {author} {\bibfnamefont {A.}~\bibnamefont {Chandra}}, \bibinfo
  {author} {\bibfnamefont {A.}~\bibnamefont {Goban}}, \bibinfo {author}
  {\bibfnamefont {A.~S.}\ \bibnamefont {Sheremet}}, \bibinfo {author}
  {\bibfnamefont {D.~V.}\ \bibnamefont {Kupriyanov}}, \ and\ \bibinfo {author}
  {\bibfnamefont {J.}~\bibnamefont {Laurat}},\ }\bibfield  {title} {\emph
  {\bibinfo {title} {{Large Bragg Reflection from One-Dimensional Chains of
  Trapped Atoms Near a Nanoscale Waveguide}},\ }}\href {\doibase
  10.1103/PhysRevLett.117.133603} {\bibfield  {journal} {\bibinfo  {journal}
  {Phys. Rev. Lett.}\ }\textbf {\bibinfo {volume} {117}},\ \bibinfo {pages}
  {133603} (\bibinfo {year} {2016})}\BibitemShut {NoStop}%
\bibitem [{\citenamefont {S\o{}rensen}\ \emph {et~al.}(2016)\citenamefont
  {S\o{}rensen}, \citenamefont {B\'eguin}, \citenamefont {Kluge}, \citenamefont
  {Iakoupov}, \citenamefont {S\o{}rensen}, \citenamefont {M\"uller},
  \citenamefont {Polzik},\ and\ \citenamefont
  {Appel}}]{PhysRevLett.117.133604}%
  \BibitemOpen
  \bibfield  {author} {\bibinfo {author} {\bibfnamefont {H.~L.}\ \bibnamefont
  {S\o{}rensen}}, \bibinfo {author} {\bibfnamefont {J.-B.}\ \bibnamefont
  {B\'eguin}}, \bibinfo {author} {\bibfnamefont {K.~W.}\ \bibnamefont {Kluge}},
  \bibinfo {author} {\bibfnamefont {I.}~\bibnamefont {Iakoupov}}, \bibinfo
  {author} {\bibfnamefont {A.~S.}\ \bibnamefont {S\o{}rensen}}, \bibinfo
  {author} {\bibfnamefont {J.~H.}\ \bibnamefont {M\"uller}}, \bibinfo {author}
  {\bibfnamefont {E.~S.}\ \bibnamefont {Polzik}}, \ and\ \bibinfo {author}
  {\bibfnamefont {J.}~\bibnamefont {Appel}},\ }\bibfield  {title} {\emph
  {\bibinfo {title} {{Coherent Backscattering of Light Off One-Dimensional
  Atomic Strings}},\ }}\href {\doibase 10.1103/PhysRevLett.117.133604}
  {\bibfield  {journal} {\bibinfo  {journal} {Phys. Rev. Lett.}\ }\textbf
  {\bibinfo {volume} {117}},\ \bibinfo {pages} {133604} (\bibinfo {year}
  {2016})}\BibitemShut {NoStop}%
\bibitem [{\citenamefont {H\"ubner}\ \emph {et~al.}(1996)\citenamefont
  {H\"ubner}, \citenamefont {Kuhl}, \citenamefont {Stroucken}, \citenamefont
  {Knorr}, \citenamefont {Koch}, \citenamefont {Hey},\ and\ \citenamefont
  {Ploog}}]{Huebner1996}%
  \BibitemOpen
  \bibfield  {author} {\bibinfo {author} {\bibfnamefont {M.}~\bibnamefont
  {H\"ubner}}, \bibinfo {author} {\bibfnamefont {J.}~\bibnamefont {Kuhl}},
  \bibinfo {author} {\bibfnamefont {T.}~\bibnamefont {Stroucken}}, \bibinfo
  {author} {\bibfnamefont {A.}~\bibnamefont {Knorr}}, \bibinfo {author}
  {\bibfnamefont {S.~W.}\ \bibnamefont {Koch}}, \bibinfo {author}
  {\bibfnamefont {R.}~\bibnamefont {Hey}}, \ and\ \bibinfo {author}
  {\bibfnamefont {K.}~\bibnamefont {Ploog}},\ }\bibfield  {title} {\emph
  {\bibinfo {title} {{Collective Effects of Excitons in Multiple-Quantum-Well
  Bragg and Anti-Bragg Structures}},\ }}\href {\doibase
  10.1103/PhysRevLett.76.4199} {\bibfield  {journal} {\bibinfo  {journal}
  {Phys. Rev. Lett.}\ }\textbf {\bibinfo {volume} {76}},\ \bibinfo {pages}
  {4199} (\bibinfo {year} {1996})}\BibitemShut {NoStop}%
\bibitem [{\citenamefont {Poddubny}(2022)}]{PhysRevA.106.L031702}%
  \BibitemOpen
  \bibfield  {author} {\bibinfo {author} {\bibfnamefont {A.~N.}\ \bibnamefont
  {Poddubny}},\ }\bibfield  {title} {\emph {\bibinfo {title} {{Driven
  anti-Bragg subradiant correlations in waveguide quantum electrodynamics}},\
  }}\href {\doibase 10.1103/PhysRevA.106.L031702} {\bibfield  {journal}
  {\bibinfo  {journal} {Phys. Rev. A}\ }\textbf {\bibinfo {volume} {106}},\
  \bibinfo {pages} {L031702} (\bibinfo {year} {2022})}\BibitemShut {NoStop}%
\bibitem [{Sup()}]{SupplementalMaterial}%
  \BibitemOpen
  \href@noop {} \bibinfo {note} {Supplemental Material can be found at https://doi.org/10.1103/PhysRevLett.131.103602, with additional detail
   about the anti-Bragg atom-dimer mirror with tunable
  reflection, atomic cavity protected by anti-parity-time symmetry and atomic
  cavity QED, which includes Ref. ~\cite{zhong2019}.}\BibitemShut {Stop}%
\bibitem [{\citenamefont {Zhong}\ \emph {et~al.}(2019)\citenamefont {Zhong},
  \citenamefont {Chang}, \citenamefont {Satzinger}, \citenamefont {Chou},
  \citenamefont {Bienfait}, \citenamefont {Conner}, \citenamefont {Dumur},
  \citenamefont {Grebel}, \citenamefont {Peairs}, \citenamefont {Povey},
  \citenamefont {Schuster},\ and\ \citenamefont {Cleland}}]{zhong2019}%
  \BibitemOpen
  \bibfield  {author} {\bibinfo {author} {\bibfnamefont {Y.~P.}\ \bibnamefont
  {Zhong}}, \bibinfo {author} {\bibfnamefont {H.-S.}\ \bibnamefont {Chang}},
  \bibinfo {author} {\bibfnamefont {K.~J.}\ \bibnamefont {Satzinger}}, \bibinfo
  {author} {\bibfnamefont {M.-H.}\ \bibnamefont {Chou}}, \bibinfo {author}
  {\bibfnamefont {A.}~\bibnamefont {Bienfait}}, \bibinfo {author}
  {\bibfnamefont {C.~R.}\ \bibnamefont {Conner}}, \bibinfo {author}
  {\bibfnamefont {{\'E}.}~\bibnamefont {Dumur}}, \bibinfo {author}
  {\bibfnamefont {J.}~\bibnamefont {Grebel}}, \bibinfo {author} {\bibfnamefont
  {G.~A.}\ \bibnamefont {Peairs}}, \bibinfo {author} {\bibfnamefont {R.~G.}\
  \bibnamefont {Povey}}, \bibinfo {author} {\bibfnamefont {D.~I.}\ \bibnamefont
  {Schuster}}, \ and\ \bibinfo {author} {\bibfnamefont {A.~N.}\ \bibnamefont
  {Cleland}},\ }\bibfield  {title} {\emph {\bibinfo {title} {{Violating Bell's
  inequality with remotely connected superconducting qubits}},\ }}\href
  {\doibase 10.1038/s41567-019-0507-7} {\bibfield  {journal} {\bibinfo
  {journal} {Nat. Phys.}\ }\textbf {\bibinfo {volume} {15}},\ \bibinfo {pages}
  {741} (\bibinfo {year} {2019})}\BibitemShut {NoStop}%
\bibitem [{\citenamefont {Le~Kien}\ \emph {et~al.}(2005)\citenamefont
  {Le~Kien}, \citenamefont {Gupta}, \citenamefont {Nayak},\ and\ \citenamefont
  {Hakuta}}]{PhysRevA.72.063815}%
  \BibitemOpen
  \bibfield  {author} {\bibinfo {author} {\bibfnamefont {F.}~\bibnamefont
  {Le~Kien}}, \bibinfo {author} {\bibfnamefont {S.~D.}\ \bibnamefont {Gupta}},
  \bibinfo {author} {\bibfnamefont {K.~P.}\ \bibnamefont {Nayak}}, \ and\
  \bibinfo {author} {\bibfnamefont {K.}~\bibnamefont {Hakuta}},\ }\bibfield
  {title} {\emph {\bibinfo {title} {{Nanofiber-mediated radiative transfer
  between two distant atoms}},\ }}\href {\doibase 10.1103/PhysRevA.72.063815}
  {\bibfield  {journal} {\bibinfo  {journal} {Phys. Rev. A}\ }\textbf {\bibinfo
  {volume} {72}},\ \bibinfo {pages} {063815} (\bibinfo {year}
  {2005})}\BibitemShut {NoStop}%
\bibitem [{\citenamefont {Gonzalez-Tudela}\ \emph {et~al.}(2011)\citenamefont
  {Gonzalez-Tudela}, \citenamefont {Martin-Cano}, \citenamefont {Moreno},
  \citenamefont {Martin-Moreno}, \citenamefont {Tejedor},\ and\ \citenamefont
  {Garcia-Vidal}}]{PhysRevLett.106.020501}%
  \BibitemOpen
  \bibfield  {author} {\bibinfo {author} {\bibfnamefont {A.}~\bibnamefont
  {Gonzalez-Tudela}}, \bibinfo {author} {\bibfnamefont {D.}~\bibnamefont
  {Martin-Cano}}, \bibinfo {author} {\bibfnamefont {E.}~\bibnamefont {Moreno}},
  \bibinfo {author} {\bibfnamefont {L.}~\bibnamefont {Martin-Moreno}}, \bibinfo
  {author} {\bibfnamefont {C.}~\bibnamefont {Tejedor}}, \ and\ \bibinfo
  {author} {\bibfnamefont {F.~J.}\ \bibnamefont {Garcia-Vidal}},\ }\bibfield
  {title} {\emph {\bibinfo {title} {{Entanglement of Two Qubits Mediated by
  One-Dimensional Plasmonic Waveguides}},\ }}\href {\doibase
  10.1103/PhysRevLett.106.020501} {\bibfield  {journal} {\bibinfo  {journal}
  {Phys. Rev. Lett.}\ }\textbf {\bibinfo {volume} {106}},\ \bibinfo {pages}
  {020501} (\bibinfo {year} {2011})}\BibitemShut {NoStop}%
\bibitem [{\citenamefont {D\'{\i}az-Camacho}\ \emph {et~al.}(2015)\citenamefont
  {D\'{\i}az-Camacho}, \citenamefont {Porras},\ and\ \citenamefont
  {Garc\'{\i}a-Ripoll}}]{PhysRevA.91.063828}%
  \BibitemOpen
  \bibfield  {author} {\bibinfo {author} {\bibfnamefont {G.}~\bibnamefont
  {D\'{\i}az-Camacho}}, \bibinfo {author} {\bibfnamefont {D.}~\bibnamefont
  {Porras}}, \ and\ \bibinfo {author} {\bibfnamefont {J.~J.}\ \bibnamefont
  {Garc\'{\i}a-Ripoll}},\ }\bibfield  {title} {\emph {\bibinfo {title}
  {{Photon-mediated qubit interactions in one-dimensional discrete and
  continuous models}},\ }}\href {\doibase 10.1103/PhysRevA.91.063828}
  {\bibfield  {journal} {\bibinfo  {journal} {Phys. Rev. A}\ }\textbf {\bibinfo
  {volume} {91}},\ \bibinfo {pages} {063828} (\bibinfo {year}
  {2015})}\BibitemShut {NoStop}%
\bibitem [{\citenamefont {Shi}\ \emph {et~al.}(2015)\citenamefont {Shi},
  \citenamefont {Chang},\ and\ \citenamefont {Cirac}}]{PhysRevA.92.053834}%
  \BibitemOpen
  \bibfield  {author} {\bibinfo {author} {\bibfnamefont {T.}~\bibnamefont
  {Shi}}, \bibinfo {author} {\bibfnamefont {D.~E.}\ \bibnamefont {Chang}}, \
  and\ \bibinfo {author} {\bibfnamefont {J.~I.}\ \bibnamefont {Cirac}},\
  }\bibfield  {title} {\emph {\bibinfo {title} {{Multiphoton-scattering theory
  and generalized master equations}},\ }}\href {\doibase
  10.1103/PhysRevA.92.053834} {\bibfield  {journal} {\bibinfo  {journal} {Phys.
  Rev. A}\ }\textbf {\bibinfo {volume} {92}},\ \bibinfo {pages} {053834}
  (\bibinfo {year} {2015})}\BibitemShut {NoStop}%
\bibitem [{\citenamefont {Caneva}\ \emph {et~al.}(2015)\citenamefont {Caneva},
  \citenamefont {Manzoni}, \citenamefont {Shi}, \citenamefont {Douglas},
  \citenamefont {Cirac},\ and\ \citenamefont {Chang}}]{caneva2015quantum}%
  \BibitemOpen
  \bibfield  {author} {\bibinfo {author} {\bibfnamefont {T.}~\bibnamefont
  {Caneva}}, \bibinfo {author} {\bibfnamefont {M.~T.}\ \bibnamefont {Manzoni}},
  \bibinfo {author} {\bibfnamefont {T.}~\bibnamefont {Shi}}, \bibinfo {author}
  {\bibfnamefont {J.~S.}\ \bibnamefont {Douglas}}, \bibinfo {author}
  {\bibfnamefont {J.~I.}\ \bibnamefont {Cirac}}, \ and\ \bibinfo {author}
  {\bibfnamefont {D.~E.}\ \bibnamefont {Chang}},\ }\bibfield  {title} {\emph
  {\bibinfo {title} {{Quantum dynamics of propagating photons with strong
  interactions: a generalized input-output formalism}},\ }}\href {\doibase
  10.1088/1367-2630/17/11/113001} {\bibfield  {journal} {\bibinfo  {journal}
  {New J. Phys.}\ }\textbf {\bibinfo {volume} {17}},\ \bibinfo {pages} {113001}
  (\bibinfo {year} {2015})}\BibitemShut {NoStop}%
\bibitem [{\citenamefont {Greenberg}\ and\ \citenamefont
  {Shtygashev}(2015)}]{PhysRevA.92.063835}%
  \BibitemOpen
  \bibfield  {author} {\bibinfo {author} {\bibfnamefont {Y.~S.}\ \bibnamefont
  {Greenberg}}\ and\ \bibinfo {author} {\bibfnamefont {A.~A.}\ \bibnamefont
  {Shtygashev}},\ }\bibfield  {title} {\emph {\bibinfo {title} {{Non-Hermitian
  Hamiltonian approach to the microwave transmission through a one-dimensional
  qubit chain}},\ }}\href {\doibase 10.1103/PhysRevA.92.063835} {\bibfield
  {journal} {\bibinfo  {journal} {Phys. Rev. A}\ }\textbf {\bibinfo {volume}
  {92}},\ \bibinfo {pages} {063835} (\bibinfo {year} {2015})}\BibitemShut
  {NoStop}%
\bibitem [{\citenamefont {Calaj\'o}\ \emph {et~al.}(2016)\citenamefont
  {Calaj\'o}, \citenamefont {Ciccarello}, \citenamefont {Chang},\ and\
  \citenamefont {Rabl}}]{PhysRevA.93.033833}%
  \BibitemOpen
  \bibfield  {author} {\bibinfo {author} {\bibfnamefont {G.}~\bibnamefont
  {Calaj\'o}}, \bibinfo {author} {\bibfnamefont {F.}~\bibnamefont
  {Ciccarello}}, \bibinfo {author} {\bibfnamefont {D.}~\bibnamefont {Chang}}, \
  and\ \bibinfo {author} {\bibfnamefont {P.}~\bibnamefont {Rabl}},\ }\bibfield
  {title} {\emph {\bibinfo {title} {{Atom-field dressed states in slow-light
  waveguide QED}},\ }}\href {\doibase 10.1103/PhysRevA.93.033833} {\bibfield
  {journal} {\bibinfo  {journal} {Phys. Rev. A}\ }\textbf {\bibinfo {volume}
  {93}},\ \bibinfo {pages} {033833} (\bibinfo {year} {2016})}\BibitemShut
  {NoStop}%
\bibitem [{\citenamefont {Ge}\ and\ \citenamefont
  {T\"ureci}(2013)}]{PhysRevA.88.053810}%
  \BibitemOpen
  \bibfield  {author} {\bibinfo {author} {\bibfnamefont {L.}~\bibnamefont
  {Ge}}\ and\ \bibinfo {author} {\bibfnamefont {H.~E.}\ \bibnamefont
  {T\"ureci}},\ }\bibfield  {title} {\emph {\bibinfo {title} {{Antisymmetric
  $\mathcal{PT}$-photonic structures with balanced positive- and negative-index
  materials}},\ }}\href {\doibase 10.1103/PhysRevA.88.053810} {\bibfield
  {journal} {\bibinfo  {journal} {Phys. Rev. A}\ }\textbf {\bibinfo {volume}
  {88}},\ \bibinfo {pages} {053810} (\bibinfo {year} {2013})}\BibitemShut
  {NoStop}%
\bibitem [{\citenamefont {Wu}\ \emph {et~al.}(2014)\citenamefont {Wu},
  \citenamefont {Artoni},\ and\ \citenamefont
  {La~Rocca}}]{PhysRevLett.113.123004}%
  \BibitemOpen
  \bibfield  {author} {\bibinfo {author} {\bibfnamefont {J.-H.}\ \bibnamefont
  {Wu}}, \bibinfo {author} {\bibfnamefont {M.}~\bibnamefont {Artoni}}, \ and\
  \bibinfo {author} {\bibfnamefont {G.~C.}\ \bibnamefont {La~Rocca}},\
  }\bibfield  {title} {\emph {\bibinfo {title} {{Non-Hermitian Degeneracies and
  Unidirectional Reflectionless Atomic Lattices}},\ }}\href {\doibase
  10.1103/PhysRevLett.113.123004} {\bibfield  {journal} {\bibinfo  {journal}
  {Phys. Rev. Lett.}\ }\textbf {\bibinfo {volume} {113}},\ \bibinfo {pages}
  {123004} (\bibinfo {year} {2014})}\BibitemShut {NoStop}%
\bibitem [{\citenamefont {Yang}\ \emph {et~al.}(2017)\citenamefont {Yang},
  \citenamefont {Liu},\ and\ \citenamefont {You}}]{PhysRevA.96.053845}%
  \BibitemOpen
  \bibfield  {author} {\bibinfo {author} {\bibfnamefont {F.}~\bibnamefont
  {Yang}}, \bibinfo {author} {\bibfnamefont {Y.-C.}\ \bibnamefont {Liu}}, \
  and\ \bibinfo {author} {\bibfnamefont {L.}~\bibnamefont {You}},\ }\bibfield
  {title} {\emph {\bibinfo {title} {{Anti-$\mathcal{PT}$ symmetry in
  dissipatively coupled optical systems}},\ }}\href {\doibase
  10.1103/PhysRevA.96.053845} {\bibfield  {journal} {\bibinfo  {journal} {Phys.
  Rev. A}\ }\textbf {\bibinfo {volume} {96}},\ \bibinfo {pages} {053845}
  (\bibinfo {year} {2017})}\BibitemShut {NoStop}%
\bibitem [{\citenamefont {Li}\ \emph {et~al.}(2019{\natexlab{b}})\citenamefont
  {Li}, \citenamefont {Zhang}, \citenamefont {Cheng}, \citenamefont {Liu},
  \citenamefont {Wang}, \citenamefont {Yan}, \citenamefont {Lin}, \citenamefont
  {Xiao}, \citenamefont {Sun}, \citenamefont {Wang}, \citenamefont {Tang},
  \citenamefont {Xu}, \citenamefont {Li},\ and\ \citenamefont {Guo}}]{li2019}%
  \BibitemOpen
  \bibfield  {author} {\bibinfo {author} {\bibfnamefont {Q.}~\bibnamefont
  {Li}}, \bibinfo {author} {\bibfnamefont {C.-J.}\ \bibnamefont {Zhang}},
  \bibinfo {author} {\bibfnamefont {Z.-D.}\ \bibnamefont {Cheng}}, \bibinfo
  {author} {\bibfnamefont {W.-Z.}\ \bibnamefont {Liu}}, \bibinfo {author}
  {\bibfnamefont {J.-F.}\ \bibnamefont {Wang}}, \bibinfo {author}
  {\bibfnamefont {F.-F.}\ \bibnamefont {Yan}}, \bibinfo {author} {\bibfnamefont
  {Z.-H.}\ \bibnamefont {Lin}}, \bibinfo {author} {\bibfnamefont
  {Y.}~\bibnamefont {Xiao}}, \bibinfo {author} {\bibfnamefont {K.}~\bibnamefont
  {Sun}}, \bibinfo {author} {\bibfnamefont {Y.-T.}\ \bibnamefont {Wang}},
  \bibinfo {author} {\bibfnamefont {J.-S.}\ \bibnamefont {Tang}}, \bibinfo
  {author} {\bibfnamefont {J.-S.}\ \bibnamefont {Xu}}, \bibinfo {author}
  {\bibfnamefont {C.-F.}\ \bibnamefont {Li}}, \ and\ \bibinfo {author}
  {\bibfnamefont {G.-C.}\ \bibnamefont {Guo}},\ }\bibfield  {title} {\emph
  {\bibinfo {title} {{Experimental simulation of anti-parity-time symmetric
  Lorentz dynamics}},\ }}\href {\doibase 10.1364/OPTICA.6.000067} {\bibfield
  {journal} {\bibinfo  {journal} {Optica}\ }\textbf {\bibinfo {volume} {6}},\
  \bibinfo {pages} {67} (\bibinfo {year} {2019}{\natexlab{b}})}\BibitemShut
  {NoStop}%
\bibitem [{\citenamefont {Kockum}\ \emph {et~al.}(2018)\citenamefont {Kockum},
  \citenamefont {Johansson},\ and\ \citenamefont
  {Nori}}]{PhysRevLett.120.140404}%
  \BibitemOpen
  \bibfield  {author} {\bibinfo {author} {\bibfnamefont {A.~F.}\ \bibnamefont
  {Kockum}}, \bibinfo {author} {\bibfnamefont {G.}~\bibnamefont {Johansson}}, \
  and\ \bibinfo {author} {\bibfnamefont {F.}~\bibnamefont {Nori}},\ }\bibfield
  {title} {\emph {\bibinfo {title} {{Decoherence-Free Interaction between Giant
  Atoms in Waveguide Quantum Electrodynamics}},\ }}\href {\doibase
  10.1103/PhysRevLett.120.140404} {\bibfield  {journal} {\bibinfo  {journal}
  {Phys. Rev. Lett.}\ }\textbf {\bibinfo {volume} {120}},\ \bibinfo {pages}
  {140404} (\bibinfo {year} {2018})}\BibitemShut {NoStop}%
\bibitem [{\citenamefont {Kannan}\ \emph {et~al.}(2020)\citenamefont {Kannan},
  \citenamefont {Ruckriegel}, \citenamefont {Campbell}, \citenamefont
  {Frisk~Kockum}, \citenamefont {Braum{\"u}ller}, \citenamefont {Kim},
  \citenamefont {Kjaergaard}, \citenamefont {Krantz}, \citenamefont {Melville},
  \citenamefont {Niedzielski}, \citenamefont {Veps{\"a}l{\"a}inen},
  \citenamefont {Winik}, \citenamefont {Yoder}, \citenamefont {Nori},
  \citenamefont {Orlando}, \citenamefont {Gustavsson},\ and\ \citenamefont
  {Oliver}}]{kannan2020waveguide}%
  \BibitemOpen
  \bibfield  {author} {\bibinfo {author} {\bibfnamefont {B.}~\bibnamefont
  {Kannan}}, \bibinfo {author} {\bibfnamefont {M.~J.}\ \bibnamefont
  {Ruckriegel}}, \bibinfo {author} {\bibfnamefont {D.~L.}\ \bibnamefont
  {Campbell}}, \bibinfo {author} {\bibfnamefont {A. F.}~\bibnamefont
  {Kockum}}, \bibinfo {author} {\bibfnamefont {J.}~\bibnamefont
  {Braum{\"u}ller}}, \bibinfo {author} {\bibfnamefont {D.~K.}\ \bibnamefont
  {Kim}}, \bibinfo {author} {\bibfnamefont {M.}~\bibnamefont {Kjaergaard}},
  \bibinfo {author} {\bibfnamefont {P.}~\bibnamefont {Krantz}}, \bibinfo
  {author} {\bibfnamefont {A.}~\bibnamefont {Melville}}, \bibinfo {author}
  {\bibfnamefont {B.~M.}\ \bibnamefont {Niedzielski}}, \bibinfo {author}
  {\bibfnamefont {A.}~\bibnamefont {Veps{\"a}l{\"a}inen}}, \bibinfo {author}
  {\bibfnamefont {R.}~\bibnamefont {Winik}}, \bibinfo {author} {\bibfnamefont
  {J. L.}~\bibnamefont {Yoder}}, \bibinfo {author} {\bibfnamefont
  {F.}~\bibnamefont {Nori}}, \bibinfo {author} {\bibfnamefont {T.~P.}\
  \bibnamefont {Orlando}}, \bibinfo {author} {\bibfnamefont {S.}~\bibnamefont
  {Gustavsson}}, \ and\ \bibinfo {author} {\bibfnamefont {W.~D.}\ \bibnamefont
  {Oliver}},\ }\bibfield  {title} {\emph {\bibinfo {title} {{Waveguide quantum
  electrodynamics with superconducting artificial giant atoms}},\ }}\href
  {\doibase 10.1038/s41586-020-2529-9} {\bibfield  {journal} {\bibinfo
  {journal} {Nature}\ }\textbf {\bibinfo {volume} {583}},\ \bibinfo {pages}
  {775} (\bibinfo {year} {2020})}\BibitemShut {NoStop}%
\bibitem [{\citenamefont {Bender}\ and\ \citenamefont
  {Boettcher}(1998)}]{PhysRevLett.80.5243}%
  \BibitemOpen
  \bibfield  {author} {\bibinfo {author} {\bibfnamefont {C.~M.}\ \bibnamefont
  {Bender}}\ and\ \bibinfo {author} {\bibfnamefont {S.}~\bibnamefont
  {Boettcher}},\ }\bibfield  {title} {\emph {\bibinfo {title} {{Real Spectra in
  Non-Hermitian Hamiltonians Having $\mathcal{PT}$ Symmetry}},\ }}\href
  {\doibase 10.1103/PhysRevLett.80.5243} {\bibfield  {journal} {\bibinfo
  {journal} {Phys. Rev. Lett.}\ }\textbf {\bibinfo {volume} {80}},\ \bibinfo
  {pages} {5243} (\bibinfo {year} {1998})}\BibitemShut {NoStop}%
\bibitem [{\citenamefont {El-Ganainy}\ \emph {et~al.}(2018)\citenamefont
  {El-Ganainy}, \citenamefont {Makris}, \citenamefont {Khajavikhan},
  \citenamefont {Musslimani}, \citenamefont {Rotter},\ and\ \citenamefont
  {Christodoulides}}]{Ganainy2018non}%
  \BibitemOpen
  \bibfield  {author} {\bibinfo {author} {\bibfnamefont {R.}~\bibnamefont
  {El-Ganainy}}, \bibinfo {author} {\bibfnamefont {K.~G.}\ \bibnamefont
  {Makris}}, \bibinfo {author} {\bibfnamefont {M.}~\bibnamefont {Khajavikhan}},
  \bibinfo {author} {\bibfnamefont {Z.~H.}\ \bibnamefont {Musslimani}},
  \bibinfo {author} {\bibfnamefont {S.}~\bibnamefont {Rotter}}, \ and\ \bibinfo
  {author} {\bibfnamefont {D.~N.}\ \bibnamefont {Christodoulides}},\ }\bibfield
   {title} {\emph {\bibinfo {title} {{Non-Hermitian physics and PT symmetry}},\
  }}\href {\doibase 10.1038/nphys4323} {\bibfield  {journal} {\bibinfo
  {journal} {Nat. Phys.}\ }\textbf {\bibinfo {volume} {14}},\ \bibinfo {pages}
  {11} (\bibinfo {year} {2018})}\BibitemShut {NoStop}%
\bibitem [{\citenamefont {{\"O}zdemir}\ \emph {et~al.}(2019)\citenamefont
  {{\"O}zdemir}, \citenamefont {Rotter}, \citenamefont {Nori},\ and\
  \citenamefont {Yang}}]{ozdemir2019parity}%
  \BibitemOpen
  \bibfield  {author} {\bibinfo {author} {\bibfnamefont {{\c{S}}.~K.}\
  \bibnamefont {{\"O}zdemir}}, \bibinfo {author} {\bibfnamefont
  {S.}~\bibnamefont {Rotter}}, \bibinfo {author} {\bibfnamefont
  {F.}~\bibnamefont {Nori}}, \ and\ \bibinfo {author} {\bibfnamefont
  {L.}~\bibnamefont {Yang}},\ }\bibfield  {title} {\emph {\bibinfo {title}
  {{Parity-time symmetry and exceptional points in photonics}},\ }}\href
  {\doibase 10.1038/s41563-019-0304-9} {\bibfield  {journal} {\bibinfo
  {journal} {Nat. Mater.}\ }\textbf {\bibinfo {volume} {18}},\ \bibinfo {pages}
  {783} (\bibinfo {year} {2019})}\BibitemShut {NoStop}%
\bibitem [{\citenamefont {Ashida}\ \emph {et~al.}(2020)\citenamefont {Ashida},
  \citenamefont {Gong},\ and\ \citenamefont {Ueda}}]{ashida2020non}%
  \BibitemOpen
  \bibfield  {author} {\bibinfo {author} {\bibfnamefont {Y.}~\bibnamefont
  {Ashida}}, \bibinfo {author} {\bibfnamefont {Z.}~\bibnamefont {Gong}}, \ and\
  \bibinfo {author} {\bibfnamefont {M.}~\bibnamefont {Ueda}},\ }\bibfield
  {title} {\emph {\bibinfo {title} {{Non-Hermitian physics}},\ }}\href
  {\doibase 10.1080/00018732.2021.1876991} {\bibfield  {journal} {\bibinfo
  {journal} {Adv. Phys.}\ }\textbf {\bibinfo {volume} {69}},\ \bibinfo {pages}
  {249} (\bibinfo {year} {2020})}\BibitemShut {NoStop}%
\bibitem [{\citenamefont {Hood}\ \emph {et~al.}(2001)\citenamefont {Hood},
  \citenamefont {Kimble},\ and\ \citenamefont {Ye}}]{PhysRevA.64.033804}%
  \BibitemOpen
  \bibfield  {author} {\bibinfo {author} {\bibfnamefont {C.~J.}\ \bibnamefont
  {Hood}}, \bibinfo {author} {\bibfnamefont {H.~J.}\ \bibnamefont {Kimble}}, \
  and\ \bibinfo {author} {\bibfnamefont {J.}~\bibnamefont {Ye}},\ }\bibfield
  {title} {\emph {\bibinfo {title} {{Characterization of high-finesse mirrors:
  Loss, phase shifts, and mode structure in an optical cavity}},\ }}\href
  {\doibase 10.1103/PhysRevA.64.033804} {\bibfield  {journal} {\bibinfo
  {journal} {Phys. Rev. A}\ }\textbf {\bibinfo {volume} {64}},\ \bibinfo
  {pages} {033804} (\bibinfo {year} {2001})}\BibitemShut {NoStop}%
\bibitem [{\citenamefont {Vu\ifmmode \check{c}\else
  \v{c}\fi{}kovi\ifmmode~\acute{c}\else \'{c}\fi{}}\ \emph
  {et~al.}(2001)\citenamefont {Vu\ifmmode \check{c}\else
  \v{c}\fi{}kovi\ifmmode~\acute{c}\else \'{c}\fi{}}, \citenamefont
  {Lon\ifmmode~\check{c}\else \v{c}\fi{}ar}, \citenamefont {Mabuchi},\ and\
  \citenamefont {Scherer}}]{PhysRevE.65.016608}%
  \BibitemOpen
  \bibfield  {author} {\bibinfo {author} {\bibfnamefont {J.}~\bibnamefont
  {Vu\ifmmode \check{c}\else \v{c}\fi{}kovi\ifmmode~\acute{c}\else
  \'{c}\fi{}}}, \bibinfo {author} {\bibfnamefont {M.}~\bibnamefont
  {Lon\ifmmode~\check{c}\else \v{c}\fi{}ar}}, \bibinfo {author} {\bibfnamefont
  {H.}~\bibnamefont {Mabuchi}}, \ and\ \bibinfo {author} {\bibfnamefont
  {A.}~\bibnamefont {Scherer}},\ }\bibfield  {title} {\emph {\bibinfo {title}
  {{Design of photonic crystal microcavities for cavity QED}},\ }}\href
  {\doibase 10.1103/PhysRevE.65.016608} {\bibfield  {journal} {\bibinfo
  {journal} {Phys. Rev. E}\ }\textbf {\bibinfo {volume} {65}},\ \bibinfo
  {pages} {016608} (\bibinfo {year} {2001})}\BibitemShut {NoStop}%
\bibitem [{\citenamefont {Albrecht}\ \emph {et~al.}(2019)\citenamefont
  {Albrecht}, \citenamefont {Henriet}, \citenamefont {Asenjo-Garcia},
  \citenamefont {Dieterle}, \citenamefont {Painter},\ and\ \citenamefont
  {Chang}}]{albrecht2019subradiant}%
  \BibitemOpen
  \bibfield  {author} {\bibinfo {author} {\bibfnamefont {A.}~\bibnamefont
  {Albrecht}}, \bibinfo {author} {\bibfnamefont {L.}~\bibnamefont {Henriet}},
  \bibinfo {author} {\bibfnamefont {A.}~\bibnamefont {Asenjo-Garcia}}, \bibinfo
  {author} {\bibfnamefont {P.~B.}\ \bibnamefont {Dieterle}}, \bibinfo {author}
  {\bibfnamefont {O.}~\bibnamefont {Painter}}, \ and\ \bibinfo {author}
  {\bibfnamefont {D.~E.}\ \bibnamefont {Chang}},\ }\bibfield  {title} {\emph
  {\bibinfo {title} {{Subradiant states of quantum bits coupled to a
  one-dimensional waveguide}},\ }}\href {\doibase 10.1088/1367-2630/ab0134}
  {\bibfield  {journal} {\bibinfo  {journal} {New J. Phys.}\ }\textbf {\bibinfo
  {volume} {21}},\ \bibinfo {pages} {025003} (\bibinfo {year}
  {2019})}\BibitemShut {NoStop}%
\bibitem [{\citenamefont {Zhang}\ and\ \citenamefont
  {M\o{}lmer}(2019)}]{PhysRevLett.122.203605}%
  \BibitemOpen
  \bibfield  {author} {\bibinfo {author} {\bibfnamefont {Y.-X.}\ \bibnamefont
  {Zhang}}\ and\ \bibinfo {author} {\bibfnamefont {K.}~\bibnamefont
  {M\o{}lmer}},\ }\bibfield  {title} {\emph {\bibinfo {title} {{Theory of
  Subradiant States of a One-Dimensional Two-Level Atom Chain}},\ }}\href
  {\doibase 10.1103/PhysRevLett.122.203605} {\bibfield  {journal} {\bibinfo
  {journal} {Phys. Rev. Lett.}\ }\textbf {\bibinfo {volume} {122}},\ \bibinfo
  {pages} {203605} (\bibinfo {year} {2019})}\BibitemShut {NoStop}%
\bibitem [{\citenamefont {Ke}\ \emph {et~al.}(2019)\citenamefont {Ke},
  \citenamefont {Poshakinskiy}, \citenamefont {Lee}, \citenamefont {Kivshar},\
  and\ \citenamefont {Poddubny}}]{PhysRevLett.123.253601}%
  \BibitemOpen
  \bibfield  {author} {\bibinfo {author} {\bibfnamefont {Y.}~\bibnamefont
  {Ke}}, \bibinfo {author} {\bibfnamefont {A.~V.}\ \bibnamefont
  {Poshakinskiy}}, \bibinfo {author} {\bibfnamefont {C.}~\bibnamefont {Lee}},
  \bibinfo {author} {\bibfnamefont {Y.~S.}\ \bibnamefont {Kivshar}}, \ and\
  \bibinfo {author} {\bibfnamefont {A.~N.}\ \bibnamefont {Poddubny}},\
  }\bibfield  {title} {\emph {\bibinfo {title} {{Inelastic Scattering of Photon
  Pairs in Qubit Arrays with Subradiant States}},\ }}\href {\doibase
  10.1103/PhysRevLett.123.253601} {\bibfield  {journal} {\bibinfo  {journal}
  {Phys. Rev. Lett.}\ }\textbf {\bibinfo {volume} {123}},\ \bibinfo {pages}
  {253601} (\bibinfo {year} {2019})}\BibitemShut {NoStop}%
\bibitem [{\citenamefont {Nie}\ \emph {et~al.}(2021)\citenamefont {Nie},
  \citenamefont {Shi}, \citenamefont {Nori},\ and\ \citenamefont
  {Liu}}]{nie2020nonreciprocal}%
  \BibitemOpen
  \bibfield  {author} {\bibinfo {author} {\bibfnamefont {W.}~\bibnamefont
  {Nie}}, \bibinfo {author} {\bibfnamefont {T.}~\bibnamefont {Shi}}, \bibinfo
  {author} {\bibfnamefont {F.}~\bibnamefont {Nori}}, \ and\ \bibinfo {author}
  {\bibfnamefont {Y.-X.}\ \bibnamefont {Liu}},\ }\bibfield  {title} {\emph
  {\bibinfo {title} {{Topology-Enhanced Nonreciprocal Scattering and Photon
  Absorption in a Waveguide}},\ }}\href {\doibase
  10.1103/PhysRevApplied.15.044041} {\bibfield  {journal} {\bibinfo  {journal}
  {Phys. Rev. Appl.}\ }\textbf {\bibinfo {volume} {15}},\ \bibinfo {pages}
  {044041} (\bibinfo {year} {2021})}\BibitemShut {NoStop}%
\bibitem [{\citenamefont {Plankensteiner}\ \emph {et~al.}(2017)\citenamefont
  {Plankensteiner}, \citenamefont {Sommer}, \citenamefont {Ritsch},\ and\
  \citenamefont {Genes}}]{PhysRevLett.119.093601}%
  \BibitemOpen
  \bibfield  {author} {\bibinfo {author} {\bibfnamefont {D.}~\bibnamefont
  {Plankensteiner}}, \bibinfo {author} {\bibfnamefont {C.}~\bibnamefont
  {Sommer}}, \bibinfo {author} {\bibfnamefont {H.}~\bibnamefont {Ritsch}}, \
  and\ \bibinfo {author} {\bibfnamefont {C.}~\bibnamefont {Genes}},\ }\bibfield
   {title} {\emph {\bibinfo {title} {{Cavity Antiresonance Spectroscopy of
  Dipole Coupled Subradiant Arrays}},\ }}\href {\doibase
  10.1103/PhysRevLett.119.093601} {\bibfield  {journal} {\bibinfo  {journal}
  {Phys. Rev. Lett.}\ }\textbf {\bibinfo {volume} {119}},\ \bibinfo {pages}
  {093601} (\bibinfo {year} {2017})}\BibitemShut {NoStop}%
\bibitem [{\citenamefont {Basov}\ \emph {et~al.}(2021)\citenamefont {Basov},
  \citenamefont {Asenjo-Garcia}, \citenamefont {Schuck}, \citenamefont {Zhu},\
  and\ \citenamefont {Rubio}}]{Basov2021}%
  \BibitemOpen
  \bibfield  {author} {\bibinfo {author} {\bibfnamefont {D.~N.}\ \bibnamefont
  {Basov}}, \bibinfo {author} {\bibfnamefont {A.}~\bibnamefont
  {Asenjo-Garcia}}, \bibinfo {author} {\bibfnamefont {P.~J.}\ \bibnamefont
  {Schuck}}, \bibinfo {author} {\bibfnamefont {X.}~\bibnamefont {Zhu}}, \ and\
  \bibinfo {author} {\bibfnamefont {A.}~\bibnamefont {Rubio}},\ }\bibfield
  {title} {\emph {\bibinfo {title} {Polariton panorama},\ }}\href {\doibase
  10.1515/nanoph-2020-0449} {\bibfield  {journal} {\bibinfo  {journal}
  {Nanophotonics}\ }\textbf {\bibinfo {volume} {10}},\ \bibinfo {pages} {549}
  (\bibinfo {year} {2021})}\BibitemShut {NoStop}%
\bibitem [{\citenamefont {Ghosh}\ \emph {et~al.}(2022)\citenamefont {Ghosh},
  \citenamefont {Su}, \citenamefont {Zhao}, \citenamefont {Fieramosca},
  \citenamefont {Wu}, \citenamefont {Li}, \citenamefont {Zhang}, \citenamefont
  {Li}, \citenamefont {Chen}, \citenamefont {Liew}, \citenamefont {Sanvitto},\
  and\ \citenamefont {Xiong}}]{ghosh2022microcavity}%
  \BibitemOpen
  \bibfield  {author} {\bibinfo {author} {\bibfnamefont {S.}~\bibnamefont
  {Ghosh}}, \bibinfo {author} {\bibfnamefont {R.}~\bibnamefont {Su}}, \bibinfo
  {author} {\bibfnamefont {J.}~\bibnamefont {Zhao}}, \bibinfo {author}
  {\bibfnamefont {A.}~\bibnamefont {Fieramosca}}, \bibinfo {author}
  {\bibfnamefont {J.}~\bibnamefont {Wu}}, \bibinfo {author} {\bibfnamefont
  {T.}~\bibnamefont {Li}}, \bibinfo {author} {\bibfnamefont {Q.}~\bibnamefont
  {Zhang}}, \bibinfo {author} {\bibfnamefont {F.}~\bibnamefont {Li}}, \bibinfo
  {author} {\bibfnamefont {Z.}~\bibnamefont {Chen}}, \bibinfo {author}
  {\bibfnamefont {T.}~\bibnamefont {Liew}}, \bibinfo {author} {\bibfnamefont
  {D.}~\bibnamefont {Sanvitto}}, \ and\ \bibinfo {author} {\bibfnamefont
  {Q.}~\bibnamefont {Xiong}},\ }\bibfield  {title} {\emph {\bibinfo {title}
  {{Microcavity exciton polaritons at room temperature}},\ }}\href {\doibase
  10.3788/PI.2022.R04} {\bibfield  {journal} {\bibinfo  {journal} {Photonics
  Insights}\ }\textbf {\bibinfo {volume} {1}},\ \bibinfo {pages} {R04}
  (\bibinfo {year} {2022})}\BibitemShut {NoStop}%
\bibitem [{\citenamefont {Jing}\ \emph {et~al.}(2017)\citenamefont {Jing},
  \citenamefont {{\"O}zdemir}, \citenamefont {L{\"u}},\ and\ \citenamefont
  {Nori}}]{jing2017high}%
  \BibitemOpen
  \bibfield  {author} {\bibinfo {author} {\bibfnamefont {H.}~\bibnamefont
  {Jing}}, \bibinfo {author} {\bibfnamefont {{\c{S}}.~K.}\ \bibnamefont
  {{\"O}zdemir}}, \bibinfo {author} {\bibfnamefont {H.}~\bibnamefont {L{\"u}}},
  \ and\ \bibinfo {author} {\bibfnamefont {F.}~\bibnamefont {Nori}},\
  }\bibfield  {title} {\emph {\bibinfo {title} {{High-order exceptional points
  in optomechanics}},\ }}\href {\doibase 10.1038/s41598-017-03546-7} {\bibfield
   {journal} {\bibinfo  {journal} {Sci. Rep.}\ }\textbf {\bibinfo {volume}
  {7}},\ \bibinfo {pages} {3386} (\bibinfo {year} {2017})}\BibitemShut
  {NoStop}%
\bibitem [{\citenamefont {Hennessy}\ \emph {et~al.}(2007)\citenamefont
  {Hennessy}, \citenamefont {Badolato}, \citenamefont {Winger}, \citenamefont
  {Gerace}, \citenamefont {Atat{\"u}re}, \citenamefont {Gulde}, \citenamefont
  {F{\"a}lt}, \citenamefont {Hu},\ and\ \citenamefont
  {Imamo{\u{g}}lu}}]{hennessy2007quantum}%
  \BibitemOpen
  \bibfield  {author} {\bibinfo {author} {\bibfnamefont {K.}~\bibnamefont
  {Hennessy}}, \bibinfo {author} {\bibfnamefont {A.}~\bibnamefont {Badolato}},
  \bibinfo {author} {\bibfnamefont {M.}~\bibnamefont {Winger}}, \bibinfo
  {author} {\bibfnamefont {D.}~\bibnamefont {Gerace}}, \bibinfo {author}
  {\bibfnamefont {M.}~\bibnamefont {Atat{\"u}re}}, \bibinfo {author}
  {\bibfnamefont {S.}~\bibnamefont {Gulde}}, \bibinfo {author} {\bibfnamefont
  {S.}~\bibnamefont {F{\"a}lt}}, \bibinfo {author} {\bibfnamefont {E.~L.}\
  \bibnamefont {Hu}}, \ and\ \bibinfo {author} {\bibfnamefont {A.}~\bibnamefont
  {Imamo{\u{g}}lu}},\ }\bibfield  {title} {\emph {\bibinfo {title} {{Quantum
  nature of a strongly coupled single quantum dot-cavity system}},\ }}\href
  {\doibase 10.1038/nature05586} {\bibfield  {journal} {\bibinfo  {journal}
  {Nature}\ }\textbf {\bibinfo {volume} {445}},\ \bibinfo {pages} {896}
  (\bibinfo {year} {2007})}\BibitemShut {NoStop}%
\bibitem [{\citenamefont {Limonov}\ \emph {et~al.}(2017)\citenamefont
  {Limonov}, \citenamefont {Rybin}, \citenamefont {Poddubny},\ and\
  \citenamefont {Kivshar}}]{limonov2017fano}%
  \BibitemOpen
  \bibfield  {author} {\bibinfo {author} {\bibfnamefont {M.~F.}\ \bibnamefont
  {Limonov}}, \bibinfo {author} {\bibfnamefont {M.~V.}\ \bibnamefont {Rybin}},
  \bibinfo {author} {\bibfnamefont {A.~N.}\ \bibnamefont {Poddubny}}, \ and\
  \bibinfo {author} {\bibfnamefont {Y.~S.}\ \bibnamefont {Kivshar}},\
  }\bibfield  {title} {\emph {\bibinfo {title} {{Fano resonances in
  photonics}},\ }}\href {\doibase 10.1038/nphoton.2017.142} {\bibfield
  {journal} {\bibinfo  {journal} {Nature Photon.}\ }\textbf {\bibinfo {volume}
  {11}},\ \bibinfo {pages} {543} (\bibinfo {year} {2017})}\BibitemShut
  {NoStop}%
\bibitem [{\citenamefont {Scully}(2015)}]{PhysRevLett.115.243602}%
  \BibitemOpen
  \bibfield  {author} {\bibinfo {author} {\bibfnamefont {M.~O.}\ \bibnamefont
  {Scully}},\ }\bibfield  {title} {\emph {\bibinfo {title} {{Single Photon
  Subradiance: Quantum Control of Spontaneous Emission and Ultrafast
  Readout}},\ }}\href {\doibase 10.1103/PhysRevLett.115.243602} {\bibfield
  {journal} {\bibinfo  {journal} {Phys. Rev. Lett.}\ }\textbf {\bibinfo
  {volume} {115}},\ \bibinfo {pages} {243602} (\bibinfo {year}
  {2015})}\BibitemShut {NoStop}%
\bibitem [{\citenamefont {Asenjo-Garcia}\ \emph {et~al.}(2017)\citenamefont
  {Asenjo-Garcia}, \citenamefont {Moreno-Cardoner}, \citenamefont {Albrecht},
  \citenamefont {Kimble},\ and\ \citenamefont {Chang}}]{PhysRevX.7.031024}%
  \BibitemOpen
  \bibfield  {author} {\bibinfo {author} {\bibfnamefont {A.}~\bibnamefont
  {Asenjo-Garcia}}, \bibinfo {author} {\bibfnamefont {M.}~\bibnamefont
  {Moreno-Cardoner}}, \bibinfo {author} {\bibfnamefont {A.}~\bibnamefont
  {Albrecht}}, \bibinfo {author} {\bibfnamefont {H.~J.}\ \bibnamefont
  {Kimble}}, \ and\ \bibinfo {author} {\bibfnamefont {D.~E.}\ \bibnamefont
  {Chang}},\ }\bibfield  {title} {\emph {\bibinfo {title} {{Exponential
  Improvement in Photon Storage Fidelities Using Subradiance and ``Selective
  Radiance'' in Atomic Arrays}},\ }}\href {\doibase 10.1103/PhysRevX.7.031024}
  {\bibfield  {journal} {\bibinfo  {journal} {Phys. Rev. X}\ }\textbf {\bibinfo
  {volume} {7}},\ \bibinfo {pages} {031024} (\bibinfo {year}
  {2017})}\BibitemShut {NoStop}%
\bibitem [{\citenamefont {Ferioli}\ \emph {et~al.}(2021)\citenamefont
  {Ferioli}, \citenamefont {Glicenstein}, \citenamefont {Henriet},
  \citenamefont {Ferrier-Barbut},\ and\ \citenamefont
  {Browaeys}}]{PhysRevX.11.021031}%
  \BibitemOpen
  \bibfield  {author} {\bibinfo {author} {\bibfnamefont {G.}~\bibnamefont
  {Ferioli}}, \bibinfo {author} {\bibfnamefont {A.}~\bibnamefont
  {Glicenstein}}, \bibinfo {author} {\bibfnamefont {L.}~\bibnamefont
  {Henriet}}, \bibinfo {author} {\bibfnamefont {I.}~\bibnamefont
  {Ferrier-Barbut}}, \ and\ \bibinfo {author} {\bibfnamefont {A.}~\bibnamefont
  {Browaeys}},\ }\bibfield  {title} {\emph {\bibinfo {title} {{Storage and
  Release of Subradiant Excitations in a Dense Atomic Cloud}},\ }}\href
  {\doibase 10.1103/PhysRevX.11.021031} {\bibfield  {journal} {\bibinfo
  {journal} {Phys. Rev. X}\ }\textbf {\bibinfo {volume} {11}},\ \bibinfo
  {pages} {021031} (\bibinfo {year} {2021})}\BibitemShut {NoStop}%
\bibitem [{\citenamefont {Rubies-Bigorda}\ \emph {et~al.}(2022)\citenamefont
  {Rubies-Bigorda}, \citenamefont {Walther}, \citenamefont {Patti},\ and\
  \citenamefont {Yelin}}]{PhysRevResearch.4.013110}%
  \BibitemOpen
  \bibfield  {author} {\bibinfo {author} {\bibfnamefont {O.}~\bibnamefont
  {Rubies-Bigorda}}, \bibinfo {author} {\bibfnamefont {V.}~\bibnamefont
  {Walther}}, \bibinfo {author} {\bibfnamefont {T.~L.}\ \bibnamefont {Patti}},
  \ and\ \bibinfo {author} {\bibfnamefont {S.~F.}\ \bibnamefont {Yelin}},\
  }\bibfield  {title} {\emph {\bibinfo {title} {{Photon control and coherent
  interactions via lattice dark states in atomic arrays}},\ }}\href {\doibase
  10.1103/PhysRevResearch.4.013110} {\bibfield  {journal} {\bibinfo  {journal}
  {Phys. Rev. Research}\ }\textbf {\bibinfo {volume} {4}},\ \bibinfo {pages}
  {013110} (\bibinfo {year} {2022})}\BibitemShut {NoStop}%
\bibitem [{\citenamefont {Holzinger}\ \emph {et~al.}(2022)\citenamefont
  {Holzinger}, \citenamefont {Guti\'errez-J\'auregui}, \citenamefont
  {H\"onigl-Decrinis}, \citenamefont {Kirchmair}, \citenamefont
  {Asenjo-Garcia},\ and\ \citenamefont {Ritsch}}]{PhysRevLett.129.253601}%
  \BibitemOpen
  \bibfield  {author} {\bibinfo {author} {\bibfnamefont {R.}~\bibnamefont
  {Holzinger}}, \bibinfo {author} {\bibfnamefont {R.}~\bibnamefont
  {Guti\'errez-J\'auregui}}, \bibinfo {author} {\bibfnamefont {T.}~\bibnamefont
  {H\"onigl-Decrinis}}, \bibinfo {author} {\bibfnamefont {G.}~\bibnamefont
  {Kirchmair}}, \bibinfo {author} {\bibfnamefont {A.}~\bibnamefont
  {Asenjo-Garcia}}, \ and\ \bibinfo {author} {\bibfnamefont {H.}~\bibnamefont
  {Ritsch}},\ }\bibfield  {title} {\emph {\bibinfo {title} {{Control of
  Localized Single- and Many-Body Dark States in Waveguide QED}},\ }}\href
  {\doibase 10.1103/PhysRevLett.129.253601} {\bibfield  {journal} {\bibinfo
  {journal} {Phys. Rev. Lett.}\ }\textbf {\bibinfo {volume} {129}},\ \bibinfo
  {pages} {253601} (\bibinfo {year} {2022})}\BibitemShut {NoStop}%
\end{thebibliography}
\end{document}